\documentclass[preprint, 11pt]{elsarticle}

\usepackage{adjustbox}              
\usepackage{amsmath}
\usepackage{color}
\usepackage{enumerate} 				

\usepackage{enumitem}				
\usepackage{graphicx}
\usepackage{multirow}				
\usepackage{tabularx} 				
\usepackage{threeparttable}
\usepackage{soul} 					
\usepackage{xargs}    				
\usepackage[svgnames]{xcolor}			

\usepackage[english]{babel}
\usepackage{caption}
\usepackage{listings}
\usepackage{amsthm}

\usepackage{algorithm2e}
\RestyleAlgo{ruled}
\usepackage{tikz}

\usepackage{float}
\usepackage{siunitx}
\usepackage{amsfonts}
\definecolor{Gray}{gray}{0.9}
\usepackage[acronym]{glossaries} 
\newacronym{ed}{ED}{Error Distance}
\newacronym{med}{MED}{Mean Error Distance}
\newacronym{nmed}{NMED}{Normalized Median Error Distance}
\newacronym{er}{ER}{Error Rate}
\newacronym{red}{RED}{Relative Error Distance}
\newacronym{hp}{HP}{Hewlett Packard}
\newacronym{mred}{MRED}{Median Relative Error Distance}
\newacronym{fa}{FA}{Full Adder}
\newacronym{imply}{IMPLY}{Material Implication}
\newacronym{magic}{MAGIC}{Memristor-Aided Logic}
\newacronym{lsb}{LSB}{Least Significant Bit}
\newacronym{msb}{MSB}{Most Significant Bit}
\newacronym{rca}{RCA}{Ripple Carry Adder}
\newacronym{soa}{SoA}{State of the Art}
\newacronym{imc}{IMC}{In-Memory Computing}
\newacronym{fom}{FOM}{Figure of Merit}
\newacronym{psnr}{PSNR}{Peak Signal-to-Noise Ratio}
\newacronym{ssim}{SSIM}{Structural Similarity Index Measure}
\newacronym{mssim}{MSSIM} {Mean Structural Similarity Index Measure}
\newacronym{mrl}{MRL}{Memristor Ratioed Logic}
\newacronym{cmos}{CMOS}{Complementary Metal-Oxide-Semiconductor}
\newacronym{kvd}{KVD}{Karnough-Veigh-Diagramm}
\newacronym{team}{TEAM}{ThrEshold Adaptive Memristor}
\newacronym{vteam}{VTEAM}{Voltage ThrEshold Adaptive Memristor}
\newacronym{nvm}{NVM}{Non-Volatile Memory}
\newacronym{tsmc}{TSMC}{Taiwan Semiconductor Manufacturing Company}
\newacronym{fpga}{FPGA}{Field Programmable Gate Array}
\newacronym{jart}{JART}{Jülich Aachen Resistive Switching Tools}
\newacronym{axc}{AxC}{Approximate Computing}
\newacronym{pim}{PIM}{Processing in Memory}
\newacronym{nmc}{NMC}{Near-Memory Computation}
\newacronym{pnm}{PNM}{Processing Near Memory}
\newacronym{nn}{NN}{Neural Network}
\newacronym{mac}{MAC}{Multiply–Accumulate}
\newacronym{cnn}{CNN}{Convolutional Neural Network}
\newacronym{axd}{AxD}{Approximation Degree}
\newacronym{axa}{AxA}{Approximated Adder}
\newacronym{ml}{ML}{Machine Learning}
\newacronym{pia}{PIA}{Processing-In-Array}
\newacronym{cad}{CAD}{Computer Aided Design}
\newacronym{pwl}{PWL}{Piece Wise Linear}
\newacronym{ai}{AI}{Artificial Intelligence}
\newacronym{mse}{MSE}{Mean Squared Error}
\newacronym{tmsl}{TMSL}{ Three Memristors Stateful Logic}
\newacronym{felix}{FELIX}{Fast and Energy efficient Logic in Memory}
\newacronym{sixor}{SIXOR}{Single Cycle In-Memristor XOR}
\newacronym{ga}{GA}{Genetic Algorithm}

\newacronym{pc}{PC}{program counter}
\newacronym{isa}{ISA}{Instruction Set Architecture}

\newacronym{rram}{RRAM}{Resistive RAM}
\usepackage[table]{xcolor}

\begin{document}

\begin{frontmatter}

\title{An Instruction Set Architecture for IMPLY-based Memristive Processing-in-Array}

\author[inst1]{Liam Splittgerber}
\ead{e62102646@student.tuwien.ac.at}

\author[inst1,inst2]{Fabian Seiler\corref{cor1}}
\ead{fabian.seiler@ziti.uni-heidelberg.de}

\author[inst1,inst2]{Nima TaheriNejad}
\ead{nima.taherinejad@ziti.uni-heidelberg.de}
\cortext[cor1]{Corresponding author}

\affiliation[inst1]{organization={Technische Universität Wien (TU Wien)},
            city={Vienna},
            country={Austria}}
\affiliation[inst2]{organization={Institute of Computer Engineering (ZITI), Heidelberg University},
            city={Heidelberg},
            country={Germany}}

\begin{abstract}
The push towards expanded ultra-low-power edge computing necessitates hardware capable of operating under extremely strict energy constraints. Traditional \gls{cmos} microcontrollers are fundamentally limited in this domain by the von Neumann bottleneck and by the static power leakage inherent to volatile memory. Memristive \gls{imc} offers a promising solution to these inefficiencies by unifying data storage and computation into a single non-volatile component. However, the \gls{soa} predominantly focuses on accelerators designed to be a co-processor for data-intensive computation. This leaves the prospect of standalone, general-purpose \gls{imc} microcontroller architectures underexplored. This thesis proposes such an architecture tailored for ultra-low-power edge devices, alongside an instruction set closely derived from the RV32I standard. Using the IMPLY stateful logic paradigm, a complete implementation of the proposed instruction set is provided, and the novel addressing schema required to support computation in the memristive crossbar array is described as well. Then, the energy use and other circuit-level metrics of the proposed architecture are evaluated through simulation and compared against those of traditional microcontrollers. Finally, the functional viability of the design is demonstrated through an application case study, describing how the proposed design could be used in an intelligent environmental sensor node. 
\end{abstract}

\end{frontmatter}

\glsresetall

\section{Introduction} \label{section:intro}

As modern computation becomes more and more data-driven, edge computing has proliferated in an attempt to move computation closer to sources of data. Many such edge computing applications demand ultra-low-power computation, with devices operating under extremely strict energy constraints \cite{Shinde2008}. The workloads of these devices typically consist of sporadic data collection and lightweight local preprocessing, followed by extended periods of dormancy. To meet these demands, attention is increasingly focused on exploring new computing paradigms and emerging technologies that can support these strict power constraints. 

Currently, traditional microcontrollers based on \gls{cmos} technology are the predominant option for low-power edge devices. However, they struggle to improve on power performance, as they all share the underlying problem of the von Neumann bottleneck \cite{Horowitz}. Executing even the simplest computations requires moving data back and forth across a power-intensive memory bus, a process that consumes significantly more energy than the computation itself \cite{TaheriNejad2024Nano}. Additionally, \gls{cmos} microcontrollers rely on volatile memory, meaning they suffer from continuous static power leakage during even their idle state, which is where many edge devices spend much of their operational lifespan. Following the philosophy of moving computation closer to data, \gls{imc} represents an emerging approach for performing computation directly within memory, offering a potential solution to the problems faced by traditional microcontrollers \cite{TaheriNejad2024Nano}. The memristor stands out as an ideal candidate for \gls{imc} memory cells, with its capability to store data through its resistive state \cite{Chua1971MM} and perform logical operations natively \cite{Borghetti2010MemSw}. Additional attributes, such as non-volatility and a highly compact form factor, further position memristors as a promising foundation for ultra-low-power edge architectures. Several developments have been made in recent years in the realm of \gls{imc} architectures based on memristive crossbar arrays. The focus of much of this work has been on exploiting the properties of memristive crossbars to develop \gls{soa} accelerator architectures capable of high-throughput vectorized computation \cite{10.1145/3466752.3480071, Leitersdorf2022AritPIM}. 

However, little work exists proposing completely standalone memristive architectures equipped with a complete \gls{isa}. Here, we propose a standalone \gls{imc} architecture that expands on \gls{soa} concepts to provide a comprehensive, general-purpose foundation for ultra-low-power edge devices. We use a design approach aimed at adapting the standard RISC--V \gls{isa} to run natively on the proposed architecture. As RISC--V is a widely adopted standard, this approach allows for an \gls{isa} that remains close to an existing familiar ecosystem, potentially easing the adoption of the proposed architecture. We handle the challenge of mapping a fundamentally load-store instruction set onto a memristive environment by introducing a novel ``address bank'' schema. This overcomes traditional register limitations, enabling static and dynamic addressing of a large two-dimensional memory space while simultaneously exploiting the vector parallelism inherent to crossbar arrays. 

This thesis is organized into six sections. An overview of the content of the individual sections and how they interact is detailed below. After the introduction in this section, we will review the literature and provide the necessary background information for this thesis in section \ref{section:literature}. This includes an overview of memristors, IMPLY and other stateful logic paradigms, and the RISC--V \gls{isa}. In section \ref{section:methodology}, we propose the hardware architecture and methodology. We introduce the component parts of the design: the crossbar array, system memory, and control logic. Then, we detail the necessary modifications made to the RISC--V \gls{isa} to make it compatible with the proposed design. In section \ref{section:Implementation}, following the architectural constraints outlined in the previous section, we present the low-level implementation of the instruction set. We provide the specific IMPLY algorithms required to execute each operation. In section \ref{section:results}, the proposed architecture is evaluated on the circuit-level using the ATOMIC simulation framework \cite{seiler2024atomic}. Afterward, the capabilities of the proposed design are exemplified through a case study. We show how the design could be used in an environmental sensor node and validate the functional completeness and ultra-low-power capabilities of the proposed architecture. We finally conclude this thesis in section \ref{section:conclusion}, where we summarize the primary gains, reiterate the novel architectural concepts introduced, and outline potential trajectories for future research.

\section{Literature Review} \label{section:literature}

\subsection{Memristors}

In 1971, L. Chua theorized the existence of a fourth fundamental passive circuit element, often referred to as the ``missing circuit element'' \cite{Chua1971MM}. This proposition was founded on the mathematical symmetry among the four fundamental circuit variables: current \textit{i}, voltage \textit{v}, charge \textit{q}, and flux linkage \textit{$\varphi$}. Combinatorially, there are six possible pairwise relationships among these four variables. However, prior to Chua's formulation, only five of these relationships were established within classical circuit theory:

\begin{enumerate}
    \item Charge-Current: $q(t) = \int_{-\infty}^ti(\tau)d\tau$
    \item Flux-Voltage: $\varphi(t) = \int_{-\infty}^tv(\tau)d\tau$
    \item Voltage-Current (Resistor): $dv = R\cdot di$
    \item Charge-Voltage (Capacitor): $dq = C \cdot dv$
    \item Flux-Current (Inductor): $d\varphi = L \cdot di$
\end{enumerate}

\begin{figure}[h]
    \centering
    \includegraphics[width=0.5\linewidth]{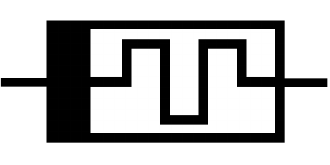}
    \caption{Circuit symbol of the memristor proposed in \cite{Chua1971MM}}
    \label{fig:memristor}
\end{figure}

The first two equations represent the fundamental time-integral definitions linking charge with current, and flux linkage with voltage. The last three equations define the relations of the three classical two-terminal passive components: the resistor, the capacitor, and the inductor, respectively. The sixth and final mathematical permutation dictates a direct constitutive relationship between flux linkage and electrical charge. Chua hypothesized that a fourth fundamental two-terminal component must exist to satisfy this theoretical gap. He termed this device the \textit{memristor}, using the standard circuit symbol shown in Figure \ref{fig:memristor}.

\begin{figure}[h]
    \centering
    \includegraphics[width=0.8\linewidth]{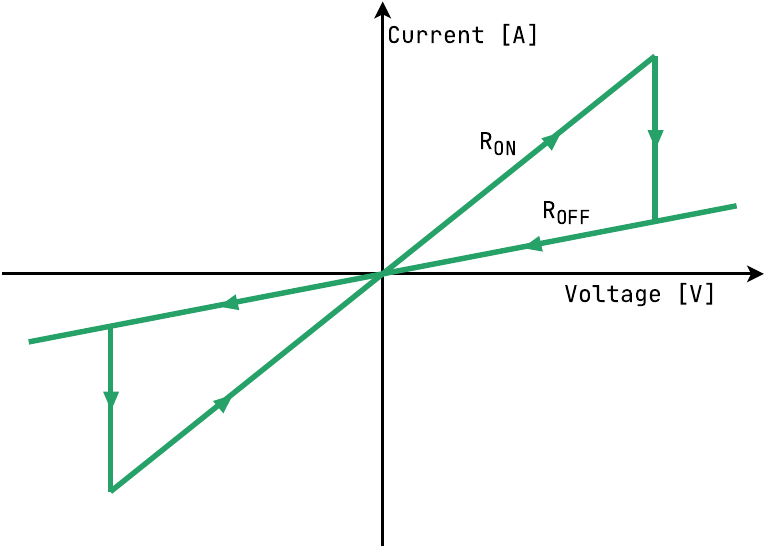}
    \caption{Theoretical ideal model of a memristor \cite{Chua1971MM, Borghetti2010MemSw}}
    \label{fig:hysteresis}
\end{figure}

The memristor is characterized by $M$, or its memristivity, which is given by the rate of change in flux as it relates to charge ($M(q) = \frac{d\varphi}{dq}$ \cite{Chua1971MM}). The units of memristivity are Ohms ($\Omega$), but crucially, it is not a static resistance; it depends on the charge, and by extension, the history of the current that has flowed through it \cite{Chua1971MM}. This dependence on the history of current flow is characterized by a hysteresis loop with two distinct stable states: a minimum $R_{on}$ and a maximum $R_{off}$ (see Figure \ref{fig:hysteresis}).

It is this bi-stable nature that makes memristors usable as a digital memory cell, where its resistive state represents its logical state. In the literature, it is convention to designate the resistive state $R_{on}$ as a logical `1' and $R_{off}$ as a logical `0' \cite{Kvatinsky2011IMPLY}. Furthermore, since the memristor's state does not change when no current flows, this memory capability is non-volatile as well. All this, along with their demonstrated compatibility with CMOS manufacturing methods \cite{Strukov2008TheMM}, has made memristors a promising emerging memory technology.

\subsection{Stateful Logic and IMPLY}

In addition to the memristor's memory capability, their ability to facilitate computation in the form of \textbf{stateful logic} is of great interest as well. In traditional \gls{cmos} logic (not stateful), the output of some logical circuitry depends only on the currently driven inputs. If the inputs stop being driven, the information is lost, i.e. there is no recoverable state. Memristors, on the other hand, allow for stateful logic. In memristive stateful logic, the physical state of a memristor dictates its logical state, and driving a collection of memristors in a specific way can update their states according to some logical combination of their initial states \cite{Borghetti2010MemSw}. The memristors act as both the storage of a logical state and as the components of a logic gate. 

Several stateful logic paradigms have been proposed to realize this behavior, each utilizing different fundamental logical primitives. The most prominent include MAGIC, which uses a NOR primitive, and FELIX, which uses a number of different single-cycle primitives (NOR, NOT, Min, NAND, and OR) \cite{Kvatinsky2014MAGIC, Gupta2018Felix}. There is also the IMPLY paradigm, based on an implication primitive, which in conjunction with the ability to reset memristors, forms a complete logic set \cite{Borghetti2010MemSw, Kvatinsky2011IMPLY}. This work uses the IMPLY paradigm, as it has proven to be the most stable approach, making it most resilient to the non-idealities of memristors \cite{Radakovits2021BELIEVER}.

\begin{figure}[h]
\centering
\includegraphics[width=0.8\textwidth, angle=0]{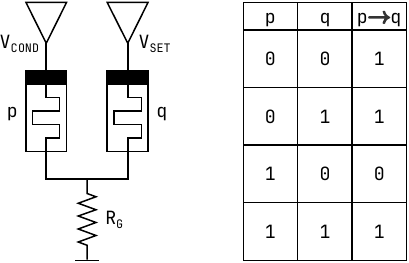}
\caption{The IMPLY logic gate using memristors}
\label{fig:imply_gate}
\end{figure}

Performing such a logical implication is done as shown in Figure \ref{fig:imply_gate}. The bottom terminals of both operand memristors $p$ and $q$ are connected to the same wire along with a resistor $\text{R}_{G}$ (where $\text{R}_{on} \ll \text{R}_{G} \ll \text{R}_{off}$). The top terminals of memristors $p$ and $q$ are connected to wires that are pulsed with voltages $\text{V}_{COND}$ and $\text{V}_{SET}$, respectively. After the voltage pulses are applied, memristor $p$ remains unchanged ($p' = p$) and the result of the logical implication overwrites the value on memristor $q$ ($q' = p \rightarrow q$). Along with the logical implication, a ``False'' operation is required as well. Resetting a memristor to its false state is done by pulsing it with the $\text{V}_{RESET}$ voltage.

\subsection{Crossbar Arrays}
\begin{figure}[h]
    \centering
    \includegraphics[width=0.8\linewidth]{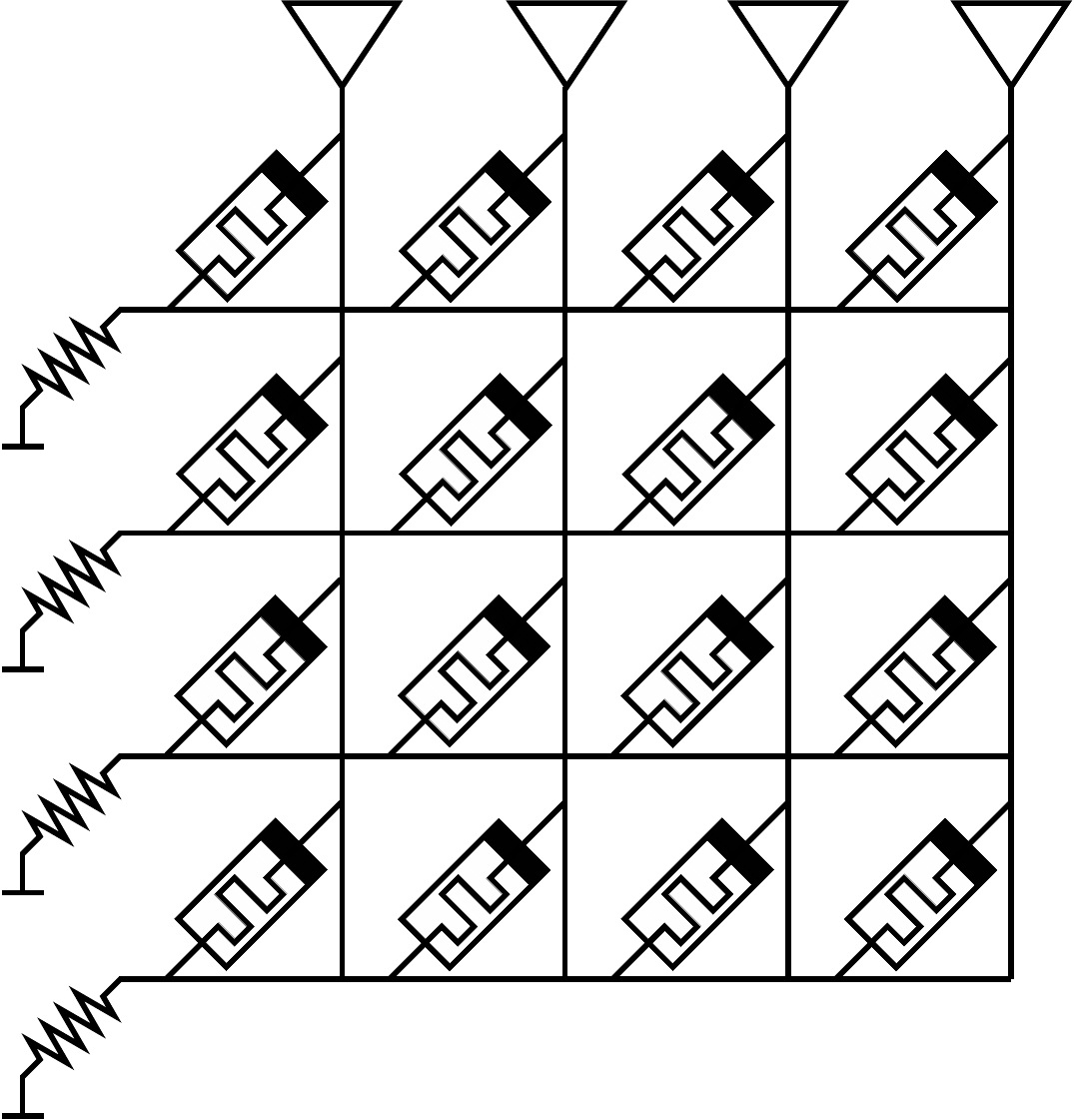}
    \caption{Memristive crossbar array for the serial topology}
    \label{fig:crossbar}
\end{figure}

In order to scale the capabilities of a single IMPLY gate, memristors can be integrated into a high-density crossbar array. This structure, shown in Figure \ref{fig:crossbar}, consists of a set of perpendicular row and column wires, with a memristor situated at each intersection. This topology is well suited for IMPLY operations because it allows the shared row wire to act as the common node for the IMPLY gate \cite{Lehtonen2009StatefulIL, Borghetti2010MemSw}. The resistor $R_G$ is connected to the end of each row, while the necessary voltages ($V_{COND}$ and $V_{SET}$) are applied with the drivers at the top of each column. Not shown in the figure are transistors that sit in series with each memristor, whose gates are connected to a set of row lines that act as a ``row select''. This is known as a One-Transistor-One-Resistor (1T1R) array. When a row is ``active'', an IMPLY operation can be applied to any arbitrary pair of memristors on that row. This is known as the serial topology. Since logic is performed bit-serially, one IMPLY step can be executed at a time within a single row.

\begin{figure}[h]
\centering
\includegraphics[width=0.5\textwidth]{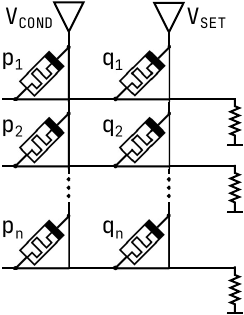}
\caption{IMPLY operation across $n$ rows}
\label{fig:parallelism}
\end{figure}

Using the serial topology of a crossbar array also provides huge parallelism capabilities. As shown in Figure \ref{fig:parallelism}, since the voltage pulses are applied to entire columns simultaneously, an IMPLY step can be performed across any number of rows in parallel without incurring any additional latency.

\subsection{Memristive \gls{imc}}
The integration of stateful logic into memristive crossbar arrays provides an ideal substrate for developing \gls{imc} architectures that avoid the von Neumann bottleneck. In traditional computer architectures, the majority of the power used is solely for transferring data between computational and memory units \cite{Horowitz, TaheriNejad2024Nano}. This fundamental bottleneck has generated a great deal of interest in computer architectures where the data never has to leave memory, and can be processed directly where the data resides. Using memristive crossbar arrays allows for just that. Data is stored in high-density memristive arrays, and stateful logic paradigms, such as IMPLY, allow for arbitrary computations within the array. Many such architectures have been proposed, for example RACER \cite{10.1145/3466752.3480071}. The RACER architecture is a scalable in-memory architecture that is designed to execute bit-pipelined bit-serial logic. In RACER, each bit of an $n$-bit data word is stored in $n$ separate small crossbar-arrays (called tiles). In order to accommodate computational patterns where data from neighboring bits is required (such as the carry bit in adders), small buffers between each tile are introduced. This allows for pipelining of operations, sacrificing as little throughput as possible. RACER is primarily designed to function as an accelerator or co-processor to aid in data-intensive applications (such as neural network inference, or image processing) rather than a standalone general-purpose processor.

Another significant development in digital \gls{imc} is the AritPIM framework \cite{Leitersdorf2022AritPIM}. AritPIM establishes a comprehensive algorithmic foundation for high throughput in-memory arithmetic. It provides a \gls{soa} suite of algorithms for all four fundamental arithmetic operations (addition, subtraction, multiplication, and division), supporting both fixed and floating-point math. AritPIM achieves high throughput by exploiting element-parallel execution across memory rows, introducing novel data-flow techniques, such as in-memory logarithmic shifting and binary-search-inspired normalization. While AritPIM provides the blueprints for high throughput computation in memristive crossbars, the performance of its algorithms are compared against a traditional GPU, indicating its intended use for accelerator hardware, not a standalone \gls{imc} architecture.

RACER and AritPIM are not alone in this regard, and research in memristive computer architectures exhibits a distinct trend: the focus is primarily on accelerator designs. These designs are implemented as co-processors that reside on a traditional system bus, intended to offload data-intensive tasks from a conventional CPU. In these designs, the CPU still handles the high-level program flow, while the memristive array is treated as a high-throughput, low-power computational slave. Little work exists proposing standalone architectures complete with a fully-equipped instruction set.

\subsection{The RISC--V Instruction Set Architecture}

Designing a general-purposed architecture involves the use of an \gls{isa} that defines the relationship between high-level software and the low-level hardware. An \gls{isa} is a specification that provides a list of operations that the processor must be able to execute and how exactly it should do so. RISC--V has emerged as the predominant \gls{isa} for academic research and novel hardware architectures due to the fact that it is an open standard and is designed for modularity and extensibility \cite{Waterman:EECS-2016-1}. The base integer instruction set, RV32I, defines a minimal, Turing-complete set of operations intended for simple architectures.

\begin{figure}[h]
\centering
\includegraphics[width=1.0\textwidth]{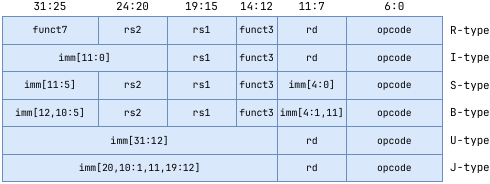}
\caption{RV32I instruction formats}
\label{fig:riscv_formats}
\end{figure}

The RISC--V architecture follows a load-store paradigm, operating on data residing in a register file \cite{Waterman:EECS-2016-1}. This means that computation is typically performed in the following pattern: first, load instructions fetch data from memory into registers; then, the CPU performs boolean or arithmetic operations on data in the register file; and finally, the resulting data is stored back to memory using store instructions. To sequence which computations happen when, control flow instructions determine the execution order of the program. All of these load, process, store, and control flow instructions belong to one of six instruction formats, as shown in Figure \ref{fig:riscv_formats}. Each instruction format has an \textbf{opcode} field, which acts as the identifier, telling the CPU what operation should be performed. Certain formats include \textbf{funct3} or \textbf{funct7} fields, which further clarify the desired operation. The six instruction formats are as follows:

\begin{enumerate}
    \item R-type (register-register): These instructions read data from two source registers (\textbf{rs1} and \textbf{rs2}), perform a boolean or arithmetic operation, and store the result to a destination register (\textbf{rd}). 
    \item I-type (immediate): These instructions source one operand from the \textbf{rs1} register, and the other directly from the 12-bit \textbf{imm} field. There are two subtypes of I-type instructions. First, there are immediate versions of many R-type instructions; these are functionally identical to the R-type version, with the exception of the second operand source. The second subtype consists of load instructions. Here, the \textbf{rs1} field holds the base address and the \textbf{imm} field represents an offset; the data at the calculated address is loaded into the \textbf{rd} register.
    \item S-type (store): S-type instructions read a word of data found in the \textbf{rs2} register, and stores it to the memory address defined by the \textbf{rs1} base address and \textbf{imm} offset.
    \item B-type (branch): These are control flow instructions that update the \gls{pc} based on a condition. The values in \textbf{rs1} and \textbf{rs2} are compared in some way, and if the comparison evaluates to 'true', the program counter is incremented by the offset found in the \textbf{imm} field. If the comparison evaluates to 'false', the program counter is incremented as normal. These instructions are used to implement if-else and loop behavior.
    \item U-type (upper immediate): These instructions have a large 20-bit \textbf{imm} field, and are employed when a 12-bit immediate value is insufficient.
    \item J-type (jump): J-type instructions are similar to B-type instructions, except they are unconditional. The value in the program counter is incremented and then stored to the \textbf{rd} register. Then, the program counter is set to the value in the \textbf{imm} field. This is the behavior of calling a function after storing the return address. 
\end{enumerate}

While the load-store paradigm that RISC--V uses has defined traditional computer architectures, it is in conflict with the principles of stateful \gls{imc}. Adapting the RISC--V \gls{isa} into a memristive environment, where data movement is eliminated and operands are not sourced from a small register file, is the central challenge addressed in the subsequent sections of this work.

\section{Methodology} \label{section:methodology}

\subsection{Proposed Architecture}

\subsubsection{Overview}

Driven by the stringent energy constraints of modern edge computing applications, the primary motivation for this design is to eliminate the power-intensive data movement and static power needs inherent to traditional von Neumann systems. To achieve this ultra-low-power computational profile, the proposed architecture is based on a memristive crossbar array, where data is stored and computation is performed. The execution of all instructions is orchestrated by a control logic unit. This unit is responsible for the complete instruction execution flow, including fetching/decoding instructions and applying the precise voltage sequences to the crossbar array to perform the desired computation. For the purposes of this thesis, the control logic is treated as a functional block, and its exact internal implementation is abstracted. The scope of this work is therefore in part to define \textit{what} the control logic must be able to do to correctly execute instructions rather than to specify exactly \textit{how} it does so.

\subsubsection{Architecture-level Model}
\begin{figure}[h]
\includegraphics[width=1\textwidth]{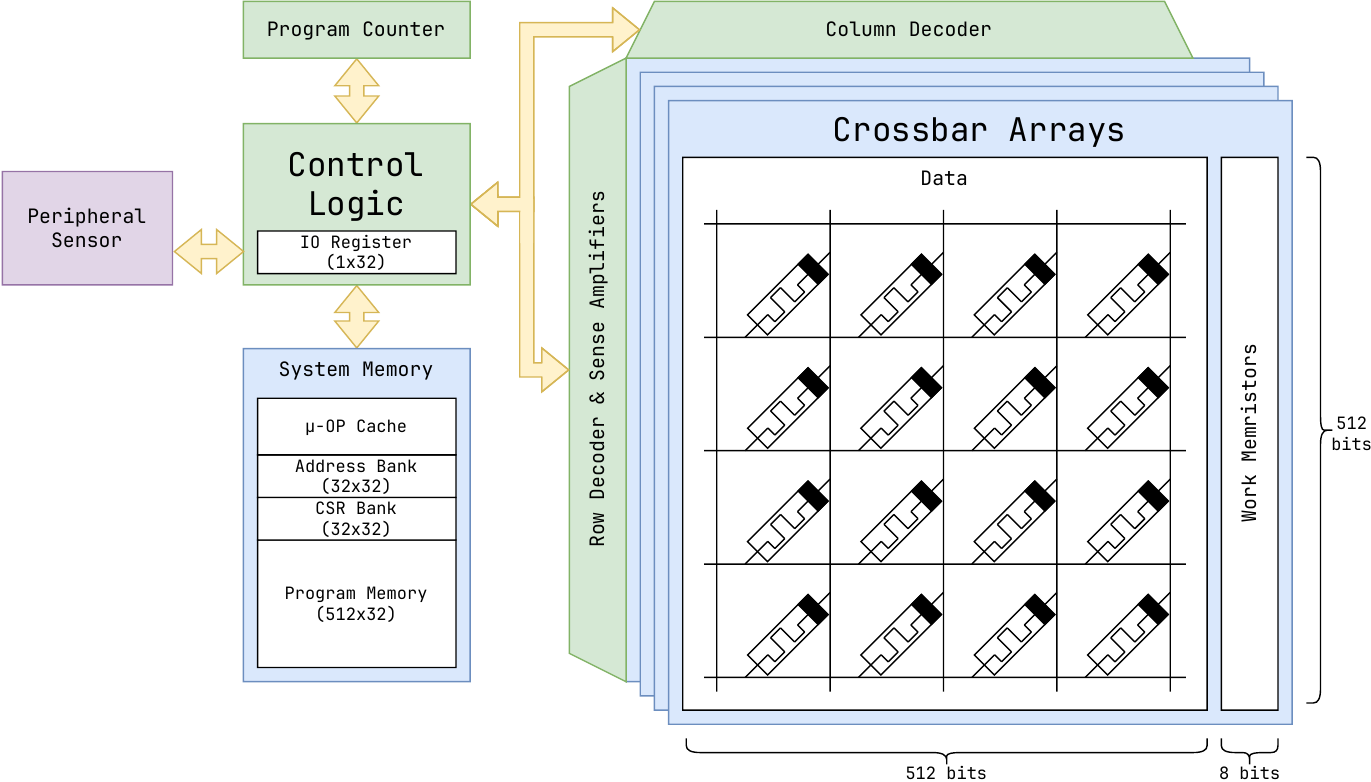}
\caption{Architecture of the crossbar array-based computer}
\label{fig:architecture_diagram}
\end{figure}

The proposed \gls{imc} architecture is shown schematically in Figure~\ref{fig:architecture_diagram}. At a high level, the system is partitioned into three primary functional blocks: the crossbar array, the control logic, and the system memory. The crossbar array serves as the combined data storage and computational medium, while the system memory holds the program and configuration data. The control logic orchestrates the entire process, interfacing with both memory structures to execute instructions. The following subsections provide a more detailed examination of each of these components and their roles.

The foundation of this in-memory architecture is the memristive \textbf{crossbar array}. Each memristor in the crossbar array holds a single bit of data by representing its logical state as the resistive state of the memristor, where a high resistance state ($\text{R}_{off}$) represents a logical `0' and a low resistance state ($\text{R}_{on}$) a logical `1'. To form multi-bit words, data is stored as contiguous blocks of memristors that are all situated on the same row. 

Computation on this data is done by composing sequences of IMPLY steps. Executing multi-step algorithms requires a place to store intermediate results, especially since the IMPLY operation is destructive by nature, overwriting one of its inputs. To address this, each row in the crossbar array is augmented with a set of dedicated work memristors. These memristors function as a scratchpad, holding the output of one logical step so it can be used as an input for another. Furthermore, they enable non-destructive computation; by first copying an operand to a work memristor, an algorithm can perform an operation without sacrificing the original input data, which may be needed for subsequent calculations.

Having established the architectural purposes of both a main data region and a dedicated set of work memristors on each row, the specific dimensions for these partitions in the proposed design can be defined. The crossbar array was chosen to have 512$\times$512 memristors for the main data region plus a 512$\times$8 section reserved for work memristors. This size was chosen as it represents a fairly large, but reasonably attainable crossbar array considering current technology scaling forecasts \cite{10.1145/3466752.3480071}. While fabricating even larger arrays is a significant research goal, practical limits are imposed by physical challenges like the resistance introduced by long wires, which can degrade the reliability of IMPLY operations.

Since the size of one array is limited, the design includes multiple crossbar arrays, all connected to the same control logic, allowing the storage capacity of the sensor node to be arbitrarily scaled. To allow for this, the control logic keeps a register with the ID of the currently active array and only executes instructions in this array. To switch to a new array when the active one is full, the control logic counts how many instructions have been executed, and when this reaches a predefined threshold, the array ID is incremented.

The system memory is an additional storage element that holds all the information that the control logic will need to execute instructions and information about the system's state. Unlike the crossbar array, no computation is performed on the data held in the system memory. Despite this, the system memory is still implemented as a memristive crossbar array. This choice is twofold: first, it leverages the same memristive fabrication process required for the main computational array, promoting a unified and potentially more cost-effective manufacturing approach. Second, by using non-volatile storage for all major memory components, the design reinforces its suitability for ultra-low-power applications. The system memory is comprised of four distinct sub-components.

\begin{enumerate}
    \item The \textbf{program memory} is a dedicated 512$\times$32 bit section responsible for storing the instructions that are to be executed. It can hold up to 512 32-bit instructions. It is given 512 rows to align with the dimensions of the main crossbar array, simplifying the manufacturing process. However, it should be noted that because the program memory is a read-only structure during execution, it is not subject to the strict voltage margins required for IMPLY operations. The voltage used to read from a memristive crossbar array ($\text{V}_{READ}$) is deliberately kept small to avoid writing to the cell, so the resistance introduced by longer wires is less of a concern. The sense amplifiers used to read from the system memory are very sensitive and can tolerate more noise. This means the program memory can theoretically be larger to accommodate more complex program code.
    \item The \textbf{address bank} holds information that the control logic uses to target the desired data in the crossbar array. Its exact function is described in more detail in Section \ref{section:addressing}.
    \item The \textbf{control and status bank} serves as an analog to the control and status registers in RISC-V. The RISC-V architecture specifies 4096 different registers for this purpose, but only a few are required for basic interrupt and trap handling \cite{Waterman:EECS-2016-1}. These registers include \textbf{mstatus}, which tracks the interrupt state and privilege levels; \textbf{mtvec}, which holds the address of the trap handler in the program memory; \textbf{mepc}, which saves the address of the instruction that caused a trap so execution can resume later; and \textbf{mcause}, which indicates the specific reason an exception or interrupt occurred. More control and status registers could be implemented, but this only has implications for exact implementation details of the control logic and is thus beyond the scope of this work.
    \item Finally, the \textbf{\textmu-OP cache} serves as a translation layer between the instruction set and the hardware. This lookup-table stores the sequences of IMPLY steps (called micro-operations) required to execute each instruction specified by the instruction set. When an instruction is to be executed, the control logic references this cache to retrieve the specific sequence of voltage pulses to apply to the crossbar array's columns to perform the desired computation.
\end{enumerate}

\begin{figure}[h]
\centering
\includegraphics[width=1\textwidth]{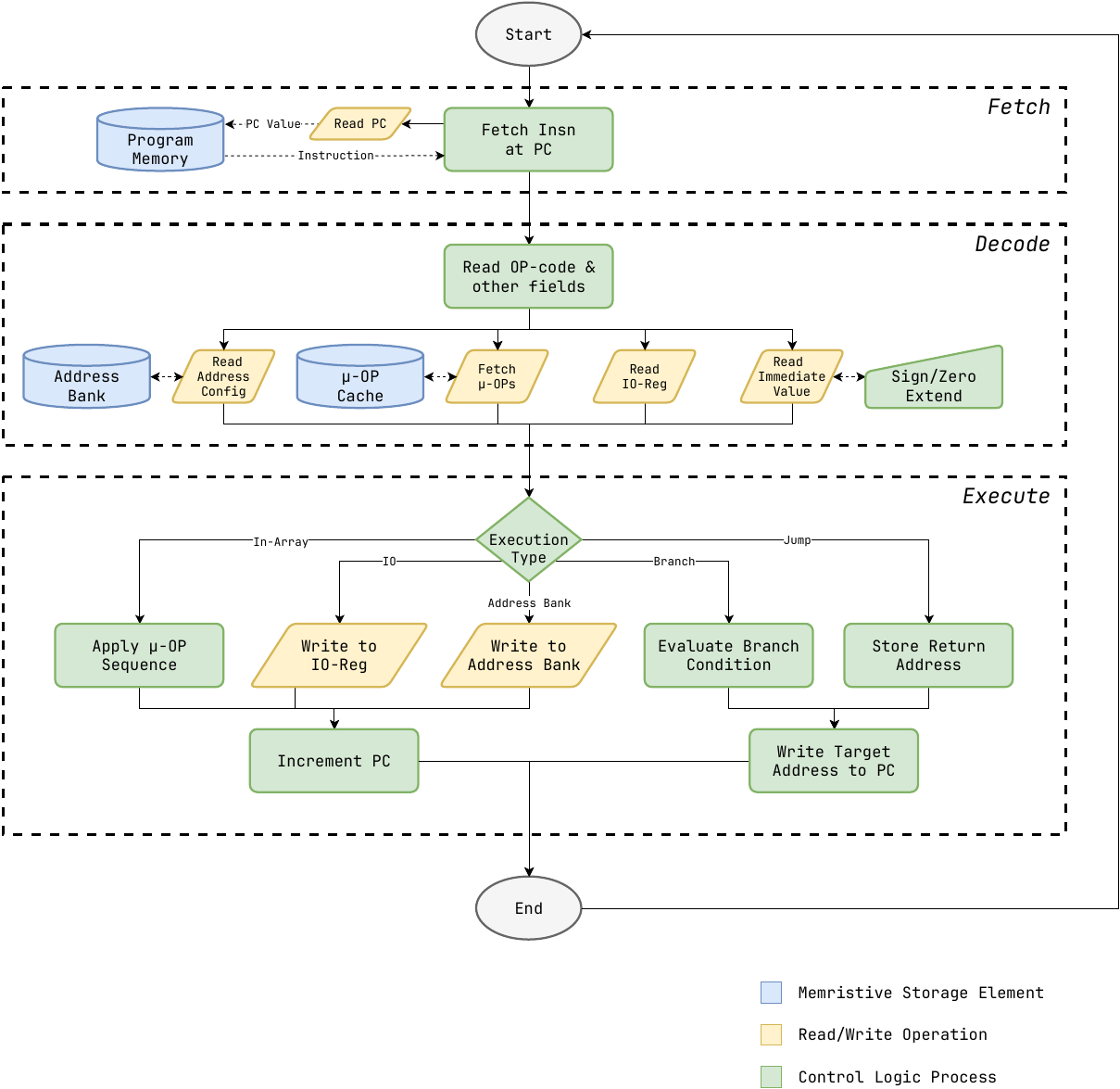}
\caption{Control flow of the architecture}
\label{fig:control_flow}
\end{figure}

The \textbf{control logic} serves as the orchestrator of the architecture, bridging the gap between the program code and the computation within the crossbar array. It is responsible for the timing, communication, and driving of all system components. At a high level, the control flow that the design uses to execute instructions is shown in Figure \ref{fig:control_flow}.

The primary section is labeled as \textbf{control logic} in Figure \ref{fig:architecture_diagram}, and it represents all the abstracted components of the control logic. It fetches instructions from the program memory, decodes them, and then performs the desired computation in the crossbar array according to the sequence of IMPLY steps found in the \textmu-OP cache. To determine which instruction is to be executed, the control logic reads from a subcomponent: the \textbf{\gls{pc}}, which simply stores the location of the next instruction to be executed in the program memory as a 32-bit value. It is implemented as a traditional volatile register instead of as a memristive storage element for two reasons. First, the \gls{pc} is incremented every clock cycle, and since no data movement is involved, it is still much more energy efficient in traditional \gls{cmos} logic. Second, non-volatility is not a required feature for the program counter, as it gets reset back to zero each time the program is run. Also contained within the control logic is the \textbf{IO register}. This register allows the design to ingest data from the external environment via the peripheral sensor. Peripherals can write to this register and trigger an interrupt, telling the design to copy its contents to the crossbar array. Additionally, this register can be used to read out the contents of the array to the external world.

Finally, the interface between the control logic and the crossbar array is handled by the \textbf{row and column decoders} and \textbf{sense amplifiers}. The decoders are implemented as demultiplexers, translating an address encoded as a binary value into a signal on the specific row and column lines corresponding to the encoded address. Integrated into the Row Decoders are the Sense Amplifiers, which read the state of memristors by applying a small voltage ($\text{V}_{READ}$) to the desired memristor, detecting the current flow, and converting the current to digital logic values. Notably, as depicted in Figure \ref{fig:architecture_diagram}, the column decoders do not extend to the work memristor portion of the array. This is to indicate that these are not directly addressable by software instructions. Still, the control logic is able to apply voltage pulses to the work memristors to facilitate intermediate steps during the IMPLY-based computations.

\subsection{Adapting the RISC-V ISA}
This architecture implements a modified version of the RV32I base integer instruction set of the RISC-V standard. The primary goal in developing the instruction set architecture for the proposed design was to adapt the RV32I instruction set to the constraints and advantages of the \gls{imc} paradigm while preserving the familiarity of the RISC-V standard as much as possible. This section first starts with an overview of the RISC-V standard, then explains and justifies the changes made. 

\subsubsection{Modifications to the RV32I Instruction Set} \label{section:isa mods}
The goal in designing the instruction set was to adhere as closely as possible to the RISC-V specification, in order to more easily adapt existing compilers for the purposes of compiling code for the design. However, some deviations are required or desirable to accommodate the \gls{imc} paradigm and its advantages. These modifications fall into three categories: the omission of memory access instructions, the introduction of new architecture-specific instructions, and fundamental changes to data-processing instructions regarding operand usage and parallelism.

\textbf{Omitted Memory Access Instructions}
The most distinct feature of this instruction set architecture is the complete omission of non-immediate load and store instructions (e.g., \texttt{lw}, \texttt{sw}, \texttt{lb}). In a traditional von Neumann architecture, these instructions shuttle data between main memory and the register file. However, in this in-memory architecture, all operands reside either within the crossbar array or as an immediate value in the program memory; there is no external RAM to load from or store to. Eliminating these instructions removes the overhead of data movement, realizing the primary energy-saving goal of the design.

The only load instructions that are kept are those that handle compile-time constants. The \textbf{Load Upper Immediate (\texttt{lui})} instruction is retained. Additionally, the \textbf{Load Immediate (\texttt{li})} pseudo-instruction, which in RISC-V is executed by involving an adder (e.g., \texttt{addi rd, x0, imm}), is implemented here as a native hardware instruction. Using an adder simply to load a constant is inefficient in a memristive environment; implementing \texttt{li} as a dedicated instruction allows the control logic to write constants directly to the crossbar array with minimal energy use.

\textbf{New Architecture-Specific Instructions}
As will be explained in detail in Section \ref{section:addressing}, in order for the control logic to know where an operand is in the crossbar array, some addressing information must already be present in the address bank. Thus, three new instructions that deal with the address bank are introduced. Note that these are designated as ``load'' instructions, but they function differently than standard RISC-V memory operations.

\begin{enumerate}
    \item \textbf{Load Address (\texttt{la}):} This instruction reads a word of data from the crossbar array and loads it to a specified slot in the address bank. This allows the program to dynamically update addresses based on values calculated in the crossbar array. The instruction needs an \textbf{rd} field to designate which slot of the address bank to store the address to. Additionally, it needs an \textbf{rs1} field to specify which word in the crossbar array should be loaded. No further fields are needed, so this instruction is encoded as an I-type instruction, where the immediate value is simply ignored.
    \item \textbf{Load Address Upper Immediate (\texttt{laui}):} Without some preexisting entries in the address bank, the control logic has no way to begin targeting operands in the crossbar array. Thus, there must be some way to load an immediate value into the address bank. Hence, \texttt{laui} is introduced. It loads a 20-bit immediate value into the upper bits of the address slot designated by the \textbf{rd} field. It is encoded as a U-type instruction to accommodate the large immediate value.
    \item \textbf{Load Address Immediate (\texttt{lai}):} Along with \texttt{laui}, an instruction is needed to populate the remaining lower 12 bits of an address slot. Hence, \texttt{lai} is introduced, which loads a 12-bit immediate value into the lower 12 bits of the address slot designated by the \textbf{rd} field. It is encoded as an I-type instruction, where the \textbf{rs1} field is ignored.
\end{enumerate}

Collectively, these three instructions provide the complete functionality required to address the crossbar array; however, two final instructions are needed to allow the system to receive and send data from the external environment. Since the architecture is designed for edge-computing applications such as sensor nodes, it requires a way to ingest data from external peripherals. Hence, \textbf{Load from IO Register (\texttt{lio})} is introduced. This instruction transfers a 32-bit value from a dedicated register (that peripherals can populate) to a location in the crossbar array, enabling the processing of externally sourced data. It is encoded as an R-type instruction, with the \textbf{rs1} field pointing to the address in the crossbar array where the value should be stored to, and all other fields being ignored. Lastly, since the data in the crossbar array must eventually be read out, the instruction \textbf{Store to IO Register (\texttt{sio})} is introduced. This instruction moves a 32-bit value from the crossbar array into the IO register, and is also encoded as an R-type instruction the same way as \texttt{lio}.
\\
\\
\textbf{Global Changes to Data-Processing Instructions}
Finally, the execution for all arithmetic and boolean instructions differs from the RISC-V standard in two ways:

\begin{enumerate}
    \item \textbf{Two-Operand Convention:} Standard RISC-V uses a three-operand format (\texttt{rd = rs1 op rs2}). In contrast, this architecture uses a two-operand, destructive write convention where the result overwrites the second input (\texttt{rs2 = rs1 op rs2}). While exceptions exist (detailed in section \ref{section:Implementation}), this approach was chosen because it is memristor and energy-efficient. Preserving both inputs of a memristive computation requires copying data to new columns, consuming additional work memristors and IMPLY steps. This also aligns with common computational patterns, such as accumulation (e.g. \texttt{sum += value}), where the original value of the destination operand is no longer required. If input preservation is absolutely necessary, the relevant input must be copied to new columns prior to the instruction.
    \item \textbf{Native Parallelism:} The design's addressing scheme leverages the massive parallelism inherent to computing in crossbar arrays with the serial topology. This vectorization comes at zero cost to latency, allowing a single instruction to process entire arrays in the same time required for a single scalar operation. 
\end{enumerate}

The next section, Section \ref{section:addressing} details a proposed addressing scheme that supports these changes.

\subsubsection{Addressing Schema} \label{section:addressing}
As previously mentioned, the proposed architecture adopts an addressing schema that is adapted from the RISC-V standard. The specific registers used for the two inputs and one output of an instruction are identified by three 5-bit fields: \textbf{rs1}, \textbf{rs2}, and \textbf{rd}. Having only 5 bits, this limits the total number of registers to $2^\text{5}=32$, as more would not be addressable without more bits. This presents a challenge for trying to adapt RISC-V for \gls{imc} purposes, as a sensible crossbar array should have far more than 32 words of storage.

To accommodate the proposed design's large address space while preserving the RISC-V instruction formats, the design introduces the \textbf{address bank}, acting as an analog to the RISC-V register file. There are 32 elements in the address bank, each with 32 bits of capacity, exactly like RISC-V's register file. An instruction can then target one of the elements of the address bank identically to a RISC-V instruction targeting an element of the register file, except instead of operating on the data in the address bank directly, the information in the address bank points to the location of the operands in the crossbar array.

\subsubsection{Proposed Address Format}
The crossbar array has 512 rows, requiring 9 bits to address, and 16 addressable words per row, requiring 4 bits to address. Given the architectural constraint that the two inputs for a given instruction must lie on the same row, addressing two inputs (2 columns and 1 row) requires 17 bits. Since the proposed architecture adopts a two-operand convention where the output overwrites the second input, these 17 bits are sufficient to address the inputs and output for a scalar operation. One of the biggest promises of \gls{imc}, however, is the vast vectorization potential, and the addressing scheme should reflect this. To provide native parallelism, each address specifies the number of rows an instruction should target beyond the base row. To clarify, if this ``Num Rows'' field is 0, the base row is still active. Additionally, should the ``Num Rows'' field contain a value larger than the amount of remaining rows, extra rows are simply ignored. Since the maximum number of rows that can be addressed is all 512 of them, this requires a further 9 bits, bringing the total number of bits for addressing to 26. This leaves 6 bits of the available 32 still unused. To make full use of the available bits, these remaining bits are used to specify a strided pattern, providing further flexibility. These bits, when interpreted as an unsigned binary number, give the interval between active rows. For instance, ``000000'' means every row in the selected range should be targeted, and ``000010'' means every fifth row should be targeted. 

\begin{figure}[h]
\includegraphics[width=1.0\textwidth]{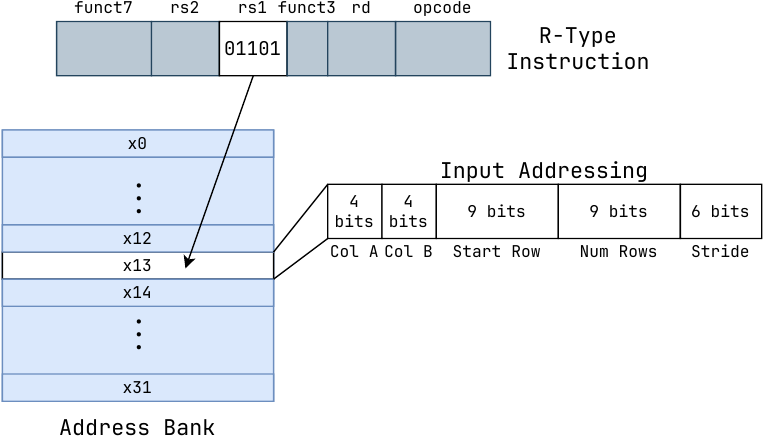}
\caption{Overview of the address bank and address format}
\label{fig:address_diagram}
\end{figure}

It should be noted that this address format fits all the information required to target the two operands in one single element of the address bank, meaning an instruction only needs one 5-bit field to target all inputs and outputs. Since most instructions have an \textbf{rs1} field, it is a logical choice to repurpose this field for addressing. In most cases, the \textbf{rs2} field is simply ignored, though specific exceptions to this rule exist and are detailed in section \ref{section:Implementation}. To summarize, Figure \ref{fig:address_diagram} depicts the function of the address bank along with the address format.

\section{Implementation} \label{section:Implementation}

This section presents a comprehensive mapping of the RV32I instruction set onto the proposed in-memory architecture. For each instruction, it provides either a detailed IMPLY-based implementation or a clear rationale for its exclusion due to fundamental incompatibility with the architecture. To provide a high-level overview before detailing the specific algorithms, Table \ref{tab:isa overview} lists the RV32I base instruction set, and classifies each instruction based on its functional category and the source of the underlying IMPLY algorithm.

Additionally, this section details the implementation of instructions beyond the standard base set. This includes common pseudo-instructions, which are implemented as native instructions to mitigate the inefficiencies of their standard definitions in a memristive environment, as well as the new custom instructions proposed in Section \ref{section:isa mods}.

\begin{table}[ph!]
    \centering
    \caption{Overview of the RV32I base set and implementation status}
    \label{tab:isa overview}   
    \resizebox{\textwidth}{!}{%
    \begin{tabular}{|c|c|c|c|}
        \hline
        Instruction & Description & Functional Type & Implementation Source\\
        \hline \hline
        \texttt{lb} & Load Byte & Load/Store & Excluded \\
        \texttt{lh} & Load Half &  & Excluded \\
        \texttt{lw} & Load Word &  & Excluded \\
        \texttt{lbu} & Load Byte Unsigned &  & Excluded \\
        \texttt{lhu} & Load Half Unsigned &  & Excluded \\
        \texttt{sb} & Store Byte &  & Excluded \\
        \texttt{sh} & Store Half &  & Excluded \\
        \texttt{sw} & Store Word &  & Excluded \\
        \texttt{lui} & Load Upper Immediate &  & Proposed \\ \hline
        \texttt{xori} & XOR Immediate & Boolean Basic & Proposed \\
        \texttt{ori} & OR Immediate &  & Proposed \\
        \texttt{andi} & AND Immediate &  & Proposed \\
        \texttt{xor} & XOR &  & Proposed \\
        \texttt{or} & OR &  & Proposed \\
        \texttt{and} & AND &  & Proposed \\ \hline
        \texttt{addi} & Add Immediate & Arithmetic & \cite{seiler2025serial} \\
        \texttt{slli} & Shift Left Logical Immediate &  & Proposed \& \cite{Leitersdorf2022AritPIM} \\
        \texttt{srli} & Shift Right Logical Immediate &  & Proposed \& \cite{Leitersdorf2022AritPIM} \\
        \texttt{srai} & Shift Right Arithmetic Immediate &  & Proposed \& \cite{Leitersdorf2022AritPIM} \\
        \texttt{add} & Add &  & \cite{seiler2025serial} \\
        \texttt{sub} & Subtract &  & \cite{9928713} \\
        \texttt{sll} & Shift Left Logical &  & Proposed \& \cite{Leitersdorf2022AritPIM} \\
        \texttt{srl} & Shift Right Logical &  & Proposed \& \cite{Leitersdorf2022AritPIM} \\
        \texttt{sra} & Shift Right Arithmetic &  & Proposed \& \cite{Leitersdorf2022AritPIM} \\
        \texttt{auipc} & Add Upper Immediate to PC &  & \cite{seiler2025serial} \\ \hline
        \texttt{slti} & Set Less Than Immediate & Comparative & Proposed \\
        \texttt{sltiu} & Set Less Than Immediate Unisgned &  & Proposed \\
        \texttt{slt} & Set Less Than &  & Proposed \\
        \texttt{sltu} & Set Less Than Unsigned &  & Proposed \\ \hline
        \texttt{beq} & Branch If Equal & Control Flow & Proposed \\
        \texttt{bne} & Branch If Not Equal &  & Proposed \\
        \texttt{blt} & Branch If Less Than &  & Proposed \\
        \texttt{bge} & Branch If Greater or Equal &  & Proposed \\
        \texttt{bltu} & Branch If Less Than Unsigned &  & Proposed \\
        \texttt{bgeu} & Branch If Greater or Equal Unsigned &  & Proposed \\
        \texttt{jalr} & Jump and Link Register &  & Proposed \\
        \texttt{jal} & Jump and Link &  & Proposed \\ \hline
        \texttt{li} & Load Immediate & Pseudo-Instructions & Proposed \\
        \texttt{mv} & Move &  & Proposed \\
        \hline
    \end{tabular}
    }        
\end{table}

\subsection{Boolean Basics}
The detailed analysis begins with the fundamental Boolean logical operations. Since the IMPLY operation, combined with a logical `0', forms a functionally complete logic system, establishing these basic logic gates is a prerequisite. These primitives serve as the building blocks upon which the more complex arithmetic and control flow algorithms presented in the following sections are constructed. In the RV32I instruction set, the native Boolean operations are limited to AND, OR, and XOR. Each of these instructions takes two 32-bit operands as inputs and performs the specified logical operation in a bit-wise fashion.

\begin{figure}[h]
\centering
\includegraphics[width=1.0\textwidth]{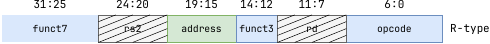}
\caption{Modified R-type encoding}
\label{fig:rtype}
\end{figure}

The instructions \texttt{and}, \texttt{or}, \texttt{xor} are all encoded as R-type instructions. This encoding has been modified to accommodate the proposed addressing scheme and is shown in Figure \ref{fig:rtype}. The \textbf{rs1} field is now used to target an entry in the address bank, and the \textbf{rs2} and \textbf{rd} fields are ignored. The functions of the other fields are kept from the RISC-V standard.

Since they all share an instruction format, the prerequisites for executing these instructions are uniform. The two source operands, denoted as A and B, occupy contiguous 32-bit blocks on the same row. Additionally, the execution of these algorithms requires one or more work memristors, which must also reside on the active row. In accordance with the architecture's two-operand convention, the final result of the operations overwrite the B operand, while the data in operand A is preserved.

\subsubsection{Instruction: \texttt{and}}
The \texttt{and} instruction performs a bit-wise AND operation on the operands $A$ and $B$. Using IMPLY logic, the AND operation can be expressed as $ab = (a \rightarrow(b \rightarrow0)) \rightarrow 0$.

The specific IMPLY algorithm used for this instruction is detailed in Table \ref{alg:and}. This sequence computes the logical AND for two single-bit operands, $a$ and $b$, utilizing one work memristor, $w_1$. To extend this operation to $n$-bit operands, the algorithm is executed sequentially for each corresponding pair of input bits. This approach allows the single work memristor $w_1$ to be reused for every iteration of the process, bringing the total number of required memristors to $2n+1$. Since each bit is processed independently, and one iteration takes 5 steps, it takes 5$n$ steps. For $n=32$, this process takes 65 memristors and 160 steps.

\begin{table}[h]
    \centering
    \caption{Proposed AND algorithm}
    \label{alg:and}   
    \begin{tabular}{|c|c|c|}
        \hline
        Steps & Operation  & Equivalent Logic  \\
        \hline \hline
        1 & $False(w_1)$ & $w_1^1=0$ \\ 
        2 & $w_1^2 = b^0 \rightarrow w_1^1$ & $w_1^2=\overline{b}$ \\ 
        3 & $w_1^3 = a^0 \rightarrow w_1^2$ & $w_1^3=\overline{ab}$  \\ 
        4 & $False(b)$ & $b^4 = 0$ \\ 
        \rowcolor{Gray} 
        5 & $b^5 = w_1^3 \rightarrow b^{4} $ & $b^5 = ab$ \\ 
        \hline
    \end{tabular}
\end{table}

\subsubsection{Instruction: \texttt{or}}
The \texttt{or} instruction performs a bit-wise logical OR on the operands $A$ and $B$. Using IMPLY logic, the OR operation can be expressed as $a+b = (a \rightarrow 0) \rightarrow b$. At a high level, the \texttt{or} instruction is performed identically to the \texttt{and} instruction, using the algorithm in Table \ref{alg:or} for each pair of input bits. For $n$ bit operands, this process takes $2n+1$ memristors and $3n$ steps. For $n=32$, this is 65 memristors and 96 steps.

\begin{table}[h]
    \centering
    \caption{Proposed OR algorithm}
    \label{alg:or}   
    \begin{tabular}{|c|c|c|}
        \hline
        Steps & Operation  & Equivalent Logic  \\
        \hline \hline
        1 & $False(w_1)$ & $w_1^1=0$ \\ 
        2 & $w_1^2 = a^0 \rightarrow w_1^1$ & $w_1^2=\overline{a}$ \\ 
        \rowcolor{Gray}
        3 & $b^3 = w_1^2 \rightarrow b^0$ & $b^3 = a + b$  \\ 
        \hline
    \end{tabular}
\end{table}

\subsubsection{Instruction: \texttt{xor}}
The \texttt{xor} instruction performs a bit-wise XOR operation on operands $A$ and $B$. Using IMPLY logic, the XOR operation can be expressed as $a \oplus b = (b \rightarrow a) \rightarrow ((a \rightarrow b) \rightarrow 0)$. At a high level, this instruction is performed almost identically to the other two boolean basic instructions, except that it requires 2 additional work memristors, or 3 in total. It performs the algorithm in Table \ref{alg:xor} for each corresponding pair of input bits. For $n$-bit operands, this instruction takes $2n+3$ memristors and $9n$ steps. For $n=32$, this is 67 memristors and 288 steps.

\begin{table}[h]
    \centering
    \caption{Proposed XOR algorithm}
    \label{alg:xor}   
    \begin{tabular}{|c|c|c|}
        \hline
        Steps & Operation  & Equivalent Logic  \\
        \hline \hline
        1 & $False(w_1, w_2, w_3)$ & $w_1^1=w_2^1=w_3^1=0$ \\ 
        2 & $w_1^2 = a^0 \rightarrow w_1^1$ & $w_1^2 = \overline{a}$ \\
        3 & $w_2^3 = b^0 \rightarrow w_2^1$ & $w_2^3 = \overline{b}$ \\
        4 & $w_3^4 = a^0 \rightarrow w_3^1$ & $w_3^4 = \overline{a}$ \\
        5 & $w_3^5 = w_2^3 \rightarrow w_3^4$ & $w_3^5 = a \rightarrow b$ \\
        6 & $w_2^6 = w_1^2 \rightarrow w_2^3$ & $w_2^6 = \overline{a} \rightarrow \overline{b}$ \\
        7 & $False(b)$ & $b^7 = 0$ \\
        8 & $b^8 = w_3^5 \rightarrow b^7$ & $b^8 = \overline{a \rightarrow b}$ \\
        \rowcolor{Gray}
        9 & $b^9 = w_2^6 \rightarrow b^8$ & $b^9 = a \oplus b$ \\
        \hline
    \end{tabular}
\end{table}

\subsubsection{Boolean Immediate Instructions}
Complementing the register-based Boolean operations, the RV32I instruction set includes immediate variants as well (\texttt{andi}, \texttt{ori}, \texttt{xori}). These instructions are functionally similar to their standard counterparts, with the distinction that the first operand is sourced directly from the instruction's 12-bit immediate field rather than from a register. 

\begin{figure}[h]
\centering
\includegraphics[width=1.0\textwidth]{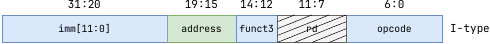}
\caption{Modified I-type encoding}
\label{fig:itype}
\end{figure}

To support this in the proposed architecture, the standard I-type instruction format is adapted as illustrated in Figure \ref{fig:itype}. Consistent with the general addressing scheme, the \textbf{rs1} field is repurposed to index the address bank, while the \textbf{rd} field is unused and ignored.

In the proposed architecture, the execution of these immediate instructions follows a two-stage process. First, the control logic reads the 12-bit immediate value from the instruction, sign-extends it to 32 bits, and writes this value directly into the crossbar array at the location designated for the first operand. This write operation is performed in two cycles: one cycle to assert the high-resistance state (logical `0') for the relevant bits, and a second cycle to assert the low-resistance state (logical `1') for the remaining bits. Once the immediate operand is present in the array, the execution proceeds identically to the corresponding standard version of the instruction.

\subsection{Arithmetic Instructions}

Building upon the Boolean primitives, this section introduces the fundamental instructions for arithmetic computation. While Boolean operations provide the basis for logic, these instructions serve as the building blocks for numerical computations, including multiplication and division. In the RV32I set, these capabilities are represented by addition (\texttt{add}), subtraction (\texttt{sub}), and a suite of shift instructions (\texttt{sll}, \texttt{srl}, \texttt{sra}). As with the Boolean instructions, these arithmetic operations use the modified R-type encoding (see Figure \ref{fig:rtype}).

The prerequisites for the arithmetic instructions mirror those of the boolean basics. The two source operands, A and B, must be stored in contiguous 32-bit blocks within the same active row. In line with the architecture's two-operand convention, the final arithmetic or shifted result is written back to the memory space originally occupied by operand $B$, while the integrity of operand $A$ is maintained.

\subsubsection{Instruction: \texttt{add}}

The \texttt{add} instruction performs the addition of two 32-bit integers using a bit-serial full-adder approach. The operation is executed by processing two input bits and a carry-in bit to generate a corresponding sum and carry-out. The process initiates at the \gls{lsb} with a carry-in of 0; the resulting carry-out is then propagated as the carry-in for the subsequent bit. In accordance with the RISC-V architecture, overflow detection for both signed and unsigned operations is not handled by the hardware, and responsibility for catching such exceptions is relegated to the programmer.

This instruction uses the full-adder algorithm shown in Table \ref{alg:fulladder} \cite{seiler2025serial}. To perform the addition in the crossbar array, the two operands must lie in two of the addressable 32-bit chunks on the same row, designated as $A$ and $B$. Additionally, four work memristors on the same row are required: $w_1$, $w_2$, and $w_3$ for storing intermediate full-adder results, and another memristor $c$ for the carry bit. Thus in total, the algorithm requires $2n+4$ memristors, which equates to 68 memristors when $n=32$. During execution, the $Sum$ output overwrites the $b$ memristor, while the $C_{out}$ overwrites the carry-in memristor $c$, allowing the same memristor to be used for the propagation of all carry bits. Upon completion, the $A$ operand is preserved, while the final sum is stored on the memristors that previously held the $B$ input. Given that the full-adder algorithm takes 20 steps, and it must be performed for each bit, the whole process takes $20n$ steps, resulting in 640 steps for $n=32$.

\begin{table}[h]
    \centering
    \caption{Full adder algorithm proposed in \cite{seiler2025serial}}
    \label{alg:fulladder}   
    \begin{tabular}{|c|c|c|}
        \hline
        Steps & Operation  & Equivalent Logic  \\
        \hline \hline
        1 & $False(w_1, w_2, w_3)$ & $w_1^1 = w_2^1 = w_3^1 = 0$ \\
        2 & $w_1^2 = a \rightarrow w_1^1$ & $w_1^2 = \overline{a}$ \\
        3 & $w_2^3 = b \rightarrow w_2^1$ & $w_2^3 = \overline{b}$ \\
        4 & $b^4 = w_1^2 \rightarrow b^0$ & $b^4 = \overline{a} \rightarrow b = X$ \\
        5 & $w_2^5 = a^0 \rightarrow w_2^3$ & $w_2^5 = a \rightarrow \overline{b} = Y$ \\
        6 & $False(w_1)$ & $w_1^6 = 0$ \\
        7 & $w_1^7 = c^0 \rightarrow w_1^2$ & $w_1^7 = \overline{c}$ \\
        8 & $c^8 = w_2^5 \rightarrow c^0$ & $c^8 = Y \rightarrow c = Z$ \\
        9 & $w_3^9 = b^4 \rightarrow w_3^1$ & $w_3^9 = \overline{X}$ \\
        10 & $w_3^{10} = w_2^5 \rightarrow w_3^9$ & $w_3^{10} = Y \rightarrow \overline{X}$ \\
        11 & $w_1^{11} = w_3^{10} \rightarrow w_1^7$ & $w_1^{11} = (Y \rightarrow \overline{X}) \rightarrow \overline{c}$ \\
        12 & $False(w_3)$ & $w_3^{12} = 0$ \\
        13 & $w_3^{13} = c^8 \rightarrow w_3^{12}$ & $w_3^{13} = \overline{Z}$ \\
        14 & $w_3^{14} = b^4 \rightarrow w_3^{13}$ & $w_3^{14} = X \rightarrow \overline{Z}$ \\
        15 & $c^{15} = b^4 \rightarrow c^8$ & $c^{15} = X \rightarrow Z$ \\
        16 & $False(b)$ & $b^{16} = 0$ \\
        17 & $b^{17} = w_1^{11} \rightarrow b^{16}$ & $b^{17} = \overline{(Y \rightarrow \overline{X}) \rightarrow \overline{c}}$ \\
        \rowcolor{Gray}
        18 & $b^{18} = c^{15} \rightarrow b^{17}$ & $b^{18} = (X \rightarrow Z) \rightarrow \overline{(Y \rightarrow \overline{X}) \rightarrow \overline{c}} = Sum$ \\
        19 & $False(c)$ & $c^{19} = 0$ \\
        \rowcolor{Gray}
        20 & $c^{20} = w_3^{14} \rightarrow c^{19}$ & $c^{20} = \overline{X \rightarrow \overline{Z}} = C_{out}$ \\
        \hline
    \end{tabular}
\end{table}

\subsubsection{Instruction: \texttt{sub}}

The \textbf{sub} instruction performs the subtraction of two 32-bit integers by using a full-subtractor. Similar to the addition process, the operation is executed bit by bit, taking two input bits and a borrow-in bit to produce a difference and a borrow-out. The execution begins at the \gls{lsb} with an initial borrow-in of 0. The resulting borrow-out from each bit is then propagated as the borrow-in for the next bit. In alignment with the RISC-V \gls{isa}, the hardware does not provide overflow or underflow detection, leaving the handling of such errors to the programmer.

The prerequisites for performing the subtraction within the crossbar array are identical to those of the \texttt{add} instruction. The two 32-bit operands must be located in contiguous blocks, $A$ and $B$, on the same active row. To facilitate the computation, four work memristors are required: $w_1$, $w_2$, and $w_3$ for intermediate state storage, and one of the borrow bit $c$. This results in a total count of $2n+4$ memristors, or 68 for a standard RV32I word length.

The subtraction is executed according to the serial full-subtractor algorithm as detailed in Table \ref{alg:fullsub} \cite{9928713}. Throughout the execution, the calculated $Diff$ bit overwrites the value in the $b$ memristor, while the $B_{out}$ bit overwrites the previous borrow-in. The allows the borrow bit to be reused for each bit, similarly to the \texttt{add} instruction. Upon completion of the instruction, the original minuend in operand A remains intact, while the final difference is stored where the $B$ operand previously was. Identically to the addition, this instruction takes $20n$ steps, or 640 steps for $n=32$.

\begin{table}[h]
    \centering
    \caption{Full subtractor algorithm proposed in \cite{9928713}}
    \label{alg:fullsub}   
    \begin{tabular}{|c|c|c|}
        \hline
        Steps & Operation  & Equivalent Logic  \\
        \hline \hline
        1 & $False(w_1, w_2, w_3)$ & $w_1^1 = w_2^1 = w_3^1 = 0$ \\
        2 & $w_1^2 = a^0 \rightarrow w_1^1$ & $w_1^2 = \overline{a}$ \\
        3 & $w_2^3 = w_1^2 \rightarrow w_2^1$ & $w_2^3 = a$ \\
        4 & $w_3^4 = w_1^2 \rightarrow w_3^1$ & $w_3^4 = a$ \\
        5 & $w_3^5 = b^0 \rightarrow w_3^4$ & $w_3^5 = b \rightarrow a$ \\
        6 & $b^6 = w_2^3 \rightarrow b^0$ & $b^6 = a \rightarrow b$ \\
        7 & $False(w_1, w_2)$ & $w_1^7 = w_2^7 = 0$ \\
        8 & $w_1^8 = b^6 \rightarrow w_1^7$ & $w_1^8 = \overline{a \rightarrow b}$ \\
        9 & $w_1^9 = w_3^5 \rightarrow w_1^8$ & $w_1^9 = a \oplus b$ \\
        10 & $w_2^{10} = w_1^9 \rightarrow w_2^7$ & $w_2^{10} = \overline{a \oplus b}$ \\
        11 & $False(b)$ & $b^{11} = 0$ \\
        12 & $b^{12} = c^0 \rightarrow b^{11}$ & $b^{12} = \overline{c}$ \\
        13 & $w_1^{13} = c^0 \rightarrow w_1^9$ & $w_1^{13} = c \rightarrow (a \oplus b)$ \\
        14 & $False(c)$ & $c^{14} = 0$ \\
        15 & $c^{15} = w_1^{13} \rightarrow c^{14}$ & $c^{15} = \overline{c \rightarrow (a \oplus b)}$ \\
        \rowcolor{Gray}
        16 & $c^{16} = w_3^5 \rightarrow c^{15}$ & $c^{16} = B_{out}$ \\
        17 & $w_2^{17} = b^{12} \rightarrow w_2^{10}$ & $w_2^{17} = \overline{c} \rightarrow \overline{(a \oplus b)}$ \\
        18 & $False(b)$ & $b^{18} = 0$ \\
        19 & $b^{19} = w_2^{17} \rightarrow b^{18}$ & $b^{19} = \overline{\overline{c} \rightarrow \overline{(a \oplus b)}}$ \\
        \rowcolor{Gray}
        20 & $b^{20} = w_1^{13} \rightarrow b^{19}$ & $b^{20} = Diff$ \\
        \hline
    \end{tabular}
\end{table}

\subsubsection{Shift Instructions}

Besides addition and subtraction, the other class of arithmetic instructions that RISC-V defines are bit shift operations. These instructions take the contents of one of its source registers, and produce a copy that is shifted left or right by a number of bits equal to the value in the second source register. Since it is not sensible to shift a value by more than $n-1$ bits, only the bottom $\text{log}_2n$ bits of the second source register are considered. 

A naive approach that was first considered for shift operations in the serial IMPLY context starts by having the control logic read the second input and determining its unsigned value. This value will be called $d$ for ``shift distance''. Then the control logic can simply copy each memristor to another memristor $d$ positions away. The problem with this approach is that it is not based solely on data-flow, and its operation depends on the value $d$ that the control logic has read. This eliminates the possibility of parallelization across rows, as the control logic would have to treat each row differently based on different values of $d$.

Another approach proposed in \cite{Leitersdorf2022AritPIM} addresses these problems. It is based only on data flow by utilizing a crossbar array based 2:1 multiplexer and a logarithmic shift approach. It is important to note that their proposed shift routine inherently operates in-place, constituting a subtle deviation from the standard two-operand convention used thus far. Whereas previous instructions preserve operand $A$ and overwrite operand $B$ with the result, this algorithm concludes with the final shifted result having overwritten the $A$ operand. 

Their proposed method, described by Algorithm \ref{alg:varshift}, uses a right logical shift as an example. If $d_j=1$, the first inner for loop is responsible for shifting $a$ by $2^j$, and the second inner for loop fills the remaining bits with zeros. If $d_j=0$, both for loops leave $a$ unchanged \cite{Leitersdorf2022AritPIM}.

\begin{algorithm}[h]
    \caption{Variable shift routine proposed in \cite{Leitersdorf2022AritPIM}}
    \label{alg:varshift}
    \SetAlgoLined
    \KwData{Input: $n$ bit $a$, $\text{log}_2n$ bit $d$, in a single row. Output: $n$ bit $a \gg d$, shifted in place}
    \KwResult{Shift operation implemented with 2:1 multiplexer and a logarithmic shift approach}
    \For{$j=0$ to $\text{log}_2n-1$}{
        \emph{Compute $a \leftarrow \text{mux}_{d_j}(a \gg 2^j, a)$ as follows:}\\
        \For{$i=0$ to $n-2^j-1$}{
            $a_i \leftarrow \text{mux}_{d_j}(a_{i+2^j}, a_i)$\;
        }
        \For{$i=n-2^j$ to $n-1$}{
            $a_i \leftarrow \overline{d_j}a_i$\;
        }
    }
\end{algorithm}

In order to realize this in the crossbar array, a suitable multiplexer algorithm is required. In order to shift in place, the output of the multiplexer algorithm should be stored on the memristor that holds its second input, or more specifically, the input chosen when $s$, the ``select'' input, is zero. The $a$ and $s$ inputs must be preserved, as they are used in a future iteration of the shift routine. The specified multiplexing operation follows the format of $a = \text{mux}_s(b,a)$. In boolean and IMPLY logic, a 2:1 multiplexer can be expressed as $\text{mux}_s(b, a) = a\overline{s} + bs = (b \rightarrow (s\rightarrow 0)) \rightarrow ((a \rightarrow s) \rightarrow 0)$. 

The proposed multiplexer algorithm detailed in Table \ref{alg:multiplex} requires two work memristors, $w_1$ and $w_2$, on the same row as inputs $a$ and $b$.

\begin{table}[h]
    \centering
    \caption{Proposed multiplexer algorithm for shift instructions}
    \label{alg:multiplex}   
    \begin{tabular}{|c|c|c|}
        \hline
        Steps & Operation  & Equivalent Logic  \\
        \hline \hline
        1 & $False(w_1, w_2)$ & $w_1^1 = w_2^1 = 0$ \\
        2 & $w_1^2 = s^0 \rightarrow w_1^0$ & $w_1^2 = \overline{s}$ \\
        3 & $w_2^3 = w_1^2 \rightarrow w_2^1$ & $w_2^3 = s$ \\
        4 & $w_1^4 = b^0 \rightarrow w_1^2$ & $w_1^4 = \overline{bs}$ \\
        5 & $w_2^5 = a^0 \rightarrow w_2^3$ & $w_2^5 = a \rightarrow s$ \\
        6 & $False(a)$ & $a^6 = 0$ \\
        7 & $a^7 = w_2^5 \rightarrow a^6$ & $a^7 = \overline{a \rightarrow s} = a\overline{s}$ \\
        \rowcolor{Gray}
        8 & $a^8 = w_1^4 \rightarrow a^7$ & $a^8 = \overline{bs} \rightarrow a\overline{s} = \text{mux}_s(a,b)$ \\
        \hline
    \end{tabular}
\end{table}

\textbf{Instruction: \texttt{srl}}

The \texttt{srl} instruction shifts an operand $A$ by up to 31 positions to the right according to the bottom 5 bits of the $B$ operand. Since it is a logical shift, the remaining bits that have not been shifted are assigned the value `0'. In the crossbar array, this is performed in exactly the manner described in Algorithm \ref{alg:varshift}, using the proposed multiplexer algorithm. The second inner for-loop requires that $a_i$ receive the value $\overline{d_j}a_i$. In IMPLY logic, this can be expressed as $(a_i \rightarrow d_j) \rightarrow 0$. To do this, the algorithm shown in Table \ref{alg:auxshift} is used. The same two work memristors used in the multiplexer algorithm can be used here, leaving the prerequisites unchanged. Upon completion of the algorithm, the $d_j$ input is preserved, and $a_i$ is either unchanged if no shift took place, or is zero if a shift occurred.

\begin{table}[h]
    \centering
    \caption{Proposed auxiliary algorithm for shift instructions}
    \label{alg:auxshift}   
    \begin{tabular}{|c|c|c|}
        \hline
        Steps & Operation  & Equivalent Logic  \\
        \hline \hline
        1 & $False(w_1, w_2)$ & $w_1^1 = w_2^1 = 0$ \\
        2 & $w_1^2 = a_i^0 \rightarrow w_1^1$ & $w_1^2 = \overline{a_i}$ \\
        3 & $w_2^3 = d_j^0 \rightarrow w_2^1$ & $w_2^3 = \overline{d_j}$ \\
        4 & $w_1^4 = w_2^3 \rightarrow w_1^2$ & $w_1^4 = d_j + \overline{a_i}$ \\
        5 & $False(a_i)$ & $a_i^5 = 0$ \\
        \rowcolor{Gray}
        6 & $a_i^6 = w_1^4 \rightarrow a_i^5$ & $a_i^6 = \overline{d_j}a_i$ \\
        \hline
    \end{tabular}
\end{table}

To obtain the total memristor count, first the inputs are considered. The primary input $a$, requires $n$ memristors, and the shift distance $d$ requires $\text{log}_2n$. Both the multiplexer algorithm and the auxiliary algorithm require two work memristors, but these can be the same two work memristors, bringing the memristor count to $n+\text{log}_2n+2$. Obtaining the step count is done by examining Algorithm \ref{alg:varshift}. The outer loop runs $\text{log}_2n$ times, and on each iteration, the multiplexer algorithm is performed $n-2^j$ times, and the auxiliary algorithm is performed $2^j$ times. This can be expressed as the following sum:
\begin{equation}
    \sum_{j=0}^{\text{log}_2n-1} (n-2^j)N_{mux} + 2^jN_{aux} = (n\text{log}_2n) N_{mux} + (n-1)(N_{aux}-N_{mux})
\end{equation}
Substituting in the step counts, this results in $8n\text{log}_2n -2n+2$ steps, or 1218 when $n=32$.

\textbf{Instruction: \texttt{sll}}

\begin{algorithm}[h]
    \caption{Proposed left logical shift routine}
    \label{alg:lvarshift}
    \SetAlgoLined
    \KwData{Input: $n$ bit $a$, $\text{log}_2n$ bit $d$, in a single row. Output: $n$ bit $a \ll d$, shifted in place}
    \KwResult{Left logical shift operation implemented with 2:1 multiplexer and a logarithmic shift approach}
    \For{$j=0$ to $\text{log}_2n-1$}{
        \emph{Compute $a \leftarrow \text{mux}_{d_j}(a \ll 2^j, a)$ as follows:}\\
        \For{$i = n-1$ to $2^j$}{
            $a_i \leftarrow \text{mux}_{d_j}(a_{i-2^j}, a_i)$\;
        }
        \For{$i = 2^j-1$ to $0$}{
            $a_i \leftarrow \overline{d_j}a_i$\;
        }
    }
\end{algorithm}

Since a left logical shift is entirely symmetric to a right logical shift, the implementation of \texttt{sll} is essentially identical to \texttt{srl}, but mirrored. The same multiplexer and auxiliary algorithm are used. Algorithm \ref{alg:varshift} is adjusted, resulting in Algorithm \ref{alg:lvarshift}.

\textbf{Instruction: \texttt{sra}}

A right arithmetic shift is slightly different than a right logical shift. Instead of writing zeros into the unshifted bits, the sign bit is copied instead. The implementation of \texttt{sra} is almost identical to \texttt{srl}, with one small change. Algorithm \ref{alg:varshift} becomes Algorithm \ref{alg:ravarshift}.

\begin{algorithm}[h]
    \caption{Proposed right arithmetic shift routine}
    \label{alg:ravarshift}
    \SetAlgoLined
    \KwData{Input: $n$ bit $a$, $\text{log}_2n$ bit $d$, in a single row. Output: $n$ bit $a \gg d$, shifted in place}
    \KwResult{Shift operation implemented with 2:1 multiplexer and a logarithmic shift approach}
    \For{$j=0$ to $\text{log}_2n-1$}{
        \emph{Compute $a \leftarrow \text{mux}_{d_j}(a \gg 2^j, a)$ as follows:}\\
        \For{$i=0$ to $n-2^j-1$}{
            $a_i \leftarrow \text{mux}_{d_j}(a_{i+2^j}, a_i)$\;
        }
        \For{$i=n-2^j$ to $n-2$}{
            $a_i \leftarrow \text{mux}_{d_j}(a_{n-1}, a_i)$\;
        }
    }
\end{algorithm}

The memristor count of this instruction is the same as the other shift instructions, but the step count is slightly different, since the auxiliary algorithm is replaced by more multiplexer algorithms. This simplifies the step count to $8n\text{log}_2n$ steps, or 1280 when $n=32$.

\subsubsection{Immediate Arithmetic Instructions}
The RV32I ISA provides immediate variants for each of the previously discussed arithmetic operations, with the exception of the \textbf{sub} instruction. These immediate instructions use the modified I-type encoding, as illustrated in Figure \ref{fig:itype}. Following the execution pattern of the Boolean immediate instructions, the process initiates by loading the sign-extended immediate value into the crossbar array, and once the immediate value is initialized within the array, the arithmetic operation proceeds identically to its normal R-type counterpart. The initialization phase introduces a constant overhead of two steps: one to set the memristors representing logic '1's and a second to set those representing logic '0's. 

For the \texttt{addi} instruction, the immediate value is written to the memristors holding the first operand, whereas for the shift immediate instructions, the immediate value gets written to the second operand memristors. The reason for the deviation from convention here is that the shift instructions shift the first operand in place, rather than overwriting the second operand, like most instructions do. For this reason, it would not make sense to use the first operand for the immediate value.

\begin{figure}[h]
\centering
\includegraphics[width=1.0\textwidth]{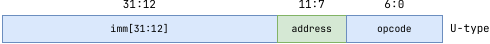}
\caption{Modified U-type encoding}
\label{fig:utype}
\end{figure}

In addition to these standard immediate operations, the RV32I set includes the \texttt{auipc} (Add Upper Immediate to PC) instruction. In the standard RISC-V specification, this instruction constructs a 32-bit offset (consisting of a 20-bit upper immediate and 12 lower zeros), adds it to the current value of the Program Counter, and stores the result in a destination register. In the proposed crossbar architecture, \texttt{auipc} functions similarly to the \texttt{addi} instruction, but utilizes the modified U-type encoding (See Figure \ref{fig:utype}) due to its larger immediate value. The difference lies in the data initialization phase. While the constructed immediate value is pasted into the first operand memristors, the control logic simultaneously pastes the current PC value into the second operand memristors. Because these writes target distinct columns, they occur in parallel and do not incur any additional steps beyond the standard two-step initialization. Once both the PC and the immediate value are present in the active row, the operation proceeds using the standard addition algorithm.

\subsection{Load Store Instructions}

In conventional von Neumann architectures, load and store instructions are required for giving the CPU access to the data held within the memory. Since the CPU cannot operate directly on data residing in main memory, values must be explicitly moved into a limited set of registers. However, in the proposed in-memory architecture, storage and computation are unified within the crossbar array. As all relevant data is resident in the array and directly addressable by the processing logic, explicit data transfer instructions become redundant. Thus, the standard RISC-V load and store instructions (\texttt{lb}, \texttt{lh}, \texttt{lw}, \texttt{lbu}, \texttt{lhu}, \texttt{sb}, \texttt{sh}, and \texttt{sw}) are not implemented.

The only class of load operations required for the proposed architecture are ``load immediate'' instructions, where the data originates from the instruction word itself rather than the memory array. The RV32I standard defines \texttt{lui} to load a 20-bit value into the upper bits of a target register. Ideally, a single instruction would load a full 32-bit immediate; however, the instruction encoding space is insufficient to contain both a 32-bit value and the necessary opcode bits. Therefore, RISC-V typically utilizes the pseudoinstruction \texttt{li} to handle 12-bit values or lower-bit insertion. In RISC-V, \texttt{li} is essentially an alias for \texttt{addi}, which adds a signed immediate to a zeroed register, and stores the result in the destination register.

In the proposed architecture, both \texttt{lui} and \texttt{li} are implemented as distinct, native instructions rather than relying on arithmetic pseudoinstructions. This design choice avoids the energy and latency overhead associated with invoking an addition for simple data initialization.

\begin{itemize}
\item \texttt{lui}: Implemented using the modified U-type encoding (Figure \ref{fig:utype}), this instruction targets the upper 20 bits of the destination block.
\item \texttt{li}: Implemented using the modified I-type encoding (Figure \ref{fig:itype}), this instruction targets the lower 12 bits.
\end{itemize}

In the crossbar array, both instructions are executed by applying $\text{V}_{SET}$ and $\text{V}_{RESET}$ signals to the desired bits. To do this, only one of the two columns in the addressing configuration is considered. To remain consistent with other immediate instructions, the ``Column A'' field is used, and the ``Column B'' field is ignored. The control logic extracts the immediate value from the instruction and performs the write in two steps: first, $\text{V}_{RESET}$ is applied to all the memristors that should hold a value of `0'; second, the specific bits corresponding to `1's in the immediate value are set by applying $\text{V}_{SET}$. This approach ensures that immediate loading is significantly faster and more energy-efficient than the equivalent arithmetic derivation used in standard RISC-V cores.

\section{Comparative Instructions}

To evaluate the relative size of two values, the RV32I instruction set includes a set of comparative operations. These instructions compare two source operands and write the resulting Boolean value to a destination register. Accordingly, standard RISC-V encodes these operations using the R-type format. The proposed architecture adapts this encoding into the modified R-type format shown in Figure \ref{fig:rtype}. Consistent with the system's two-operand policy, the boolean result of the comparison is written directly over the second source operand, replacing the original data.

\subsection{Instruction: \texttt{sltu}}
In RISC-V, the \texttt{sltu} instruction compares two 32-bit unsigned integers. It sets the destination register to 1 if the first operand is strictly less than the second, and to 0 otherwise. To achieve this in the crossbar array, first a suitable single-bit comparator algorithm is needed. This algorithm takes in two input memristors and produces two distinct outputs: a ``Less Than'' bit ($L$), which indicates if the first bit is strictly smaller than the second, and an ``Equal'' bit ($E$), which indicates if the two bits are identical. In IMPLY logic, a ``Less Than'' comparison is expressed as $\overline{b \rightarrow a}$, and an equality check is simply a NXOR. The comparator algorithm is shown in Table \ref{alg:comparator}, and requires two input memristors, $a$ and $b$, and three work memristors, all on the same row. 

\begin{table}[h]
    \centering
    \caption{Proposed single-bit comparator}
    \label{alg:comparator}   
    \begin{tabular}{|c|c|c|}
        \hline
        Steps & Operation  & Equivalent Logic  \\
        \hline \hline
        1 & $False(w_1, w_2, w_3)$ & $w_1^1=w_2^1=w_3^1=0$ \\ 
        2 & $w_1^2 = a^0 \rightarrow w_1^1$ & $w_1^2 = \overline{a}$ \\
        3 & $w_2^3 = b^0 \rightarrow w_2^1$ & $w_2^3 = \overline{b}$ \\
        4 & $w_3^4 = b^0 \rightarrow w_3^1$ & $w_3^4 = \overline{b}$ \\
        5 & $w_2^5 = w_1^2 \rightarrow w_2^3$ & $w_2^5 = b \rightarrow a$ \\
        6 & $w_1^6 = w_3^4 \rightarrow w_1^2$ & $w_1^6 = a \rightarrow b$ \\
        7 & $False(w_3)$ & $w_3^7 = 0$ \\
        8 & $w_3^8 = w_1^6 \rightarrow w_3^7$ & $w_3^8 = \overline{a \rightarrow b}$ \\
        9 & $w_3^9 = w_2^5 \rightarrow w_3^8$ & $w_3^9 = a \oplus b$ \\
        10 & $False(w_1)$ & $w_1^{10} = 0$ \\
        \rowcolor{Gray}
        11 & $w_1^{11} = w_3^9 \rightarrow w_1^{10}$ & $w_1^{11} = \overline{a \oplus b} = E$ \\
        12 & $False(b)$ & $b^{12} = 0$ \\
        \rowcolor{Gray}
        13 & $b^{13} = w_2^5 \rightarrow b^{12}$ & $b^{13} = \overline{b \rightarrow a} = L$ \\
        \hline
    \end{tabular}
\end{table}

In order to extend this to $n$ bit comparisons, additional logic is necessary. A simple approach is to first produce the $E$ and $L$ outputs for each bit. These will be called $E_i$ and $L_i$. Then, the comparison starts at the \gls{msb}, or at index $n-1$. If $L_{n-1}$ is set, it is already clear that the first input is less than the second. If $E_{n-1}$ is set, then the next most significant bit is the most relevant, and the process repeats. Formally, the logic that this process describes is:

\begin{equation}
    \sum_{i=0}^{n-1}L_i(\prod_{j=i+1}^{n-1}E_j)
\end{equation}

For clarity, each summand in the case where $n=4$ is listed.
\begin{itemize}

    \item $i=3$: $L_3$
    \item $i=2$: $E_3L_2$
    \item $i=1$: $E_3E_2L_1$
    \item $i=0$: $E_3E_2E_1L_0$
\end{itemize}

After all summands are combined with an OR operation, the comparison is complete. This approach works, but suffers from high memristor usage. Along with the inputs, at least $n$ work memristors are required to hold the $E_i$ outputs. This can be reduced to only four work memristors simply by changing the order in which operations are performed. This process is described by Algorithm \ref{alg:sltu}.

\begin{algorithm}[h]
    \caption{Proposed less-than comparison routine}
    \label{alg:sltu}
    \SetAlgoLined
    \KwData{Input: $n$ bit $a$, $n$ bit $b$, in a single row. Output: 1 bit comparison output}
    1) Produce $E_{n-1}$ and $L_{n-1}$, stored on a work memristor and the $b_{n-1}$ memristor, respectively\;
    2)
    \For{$i=n-2$ to $0$}{
        Produce $E_i$ and $L_i$, stored on a work memristor and the $b_i$ memristor, respectively\;
        Create current summand: $L_i' \leftarrow E_{i+1}L_i$\;
        Accumulate summands: $L_i'' \leftarrow L_{i+1}+L_i'$\;
        Accumulate equality product: $E_i' \leftarrow E_{i+1}E_i$\;
        \emph{The memristors holding $L_{i+1}$ and $E_{i+1}$ are now free to be overwritten}\;
    }
    3) Reset memristors $(L_{n-1} : L_1)$;
\end{algorithm} 
\vspace{8mm}
After this algorithm is done, the final result is stored on the memristor corresponding to the least significant bit of the second operand ($b_0$, now designated as $L_0$), while the entire first operand ($a$) is preserved. The total required memristor count for this instruction is $2n+4$: $2n$ for the inputs, 3 work memristors for the comparator, and 1 to hold the accumulating equality product. 

The latency of the instruction is derived from the sequence detailed in Algorithm \ref{alg:sltu}. Step 1 consists of one comparator algorithm, adding 13 steps. Then, in step 2, a for loop runs for $n-1$ iterations. Each iteration requires a comparator, two AND operations, and an OR operation, bringing the step count per iteration to 26. Finally, step 3 requires one cycle to complete. This is $26n-12$ steps in total. 


\subsection{Instruction: \texttt{slt}}
The implementation of \texttt{slt} introduces additional complexity compared to \texttt{sltu}. Since signed integers are represented in 2s complement form, the sign bit must be considered. If the sign bit in $a$ is the same as in $b$, the process for the remaining bits is identical to \texttt{sltu}. Otherwise, the sign bit is the only relevant bit. The logic that describes this process is as follows.
\begin{equation}
    \texttt{slt }a,b = \overline{E_{n-1}} \text{ } \overline{L_{n-1}}+E_{n-1}( \texttt{sltu } a[n-2:0], b[n-2:0])
\end{equation}
This form has some issues, namely that both $\overline{E_{n-1}}$ and $E_{n-1}$ are required, so an equivalent form is better suited. First, one can recognize that $\overline{p} \text{ } \overline{q}$ is equivalent to $\overline{p+q}$. This allows the first summand to be expressed as $\overline{E_{n-1}+L_{n-1}}$. Then, another simplification can be made by recognizing that $\overline{p}+q$ is identical to $p \rightarrow q$. Using this, the whole expressions becomes
\begin{equation}
    \texttt{slt }a,b = (E_{n-1} + L_{n-1}) \rightarrow E_{n-1}( \texttt{sltu } a[n-2:0], b[n-2:0])
\end{equation}
The final output is again stored on the $L_0$ memristor.

The amount of memristors used is $2n+4$. $2n$ are needed to hold the two inputs, and evaluating \texttt{sltu} for everything but the \gls{msb} requires four work memristors. The remaining logic requires less than four work memristors, making the \texttt{sltu} portion the limiting factor. To find the step count, first the \texttt{sltu} portion is considered, which takes $26n-12$ steps. Since it acts on everything but the \gls{msb}, $n$ can be replaced with $n-1$, resulting in $26n-38$ steps. Then, the comparator is run on the \gls{msb}, adding 13 more steps. The remaining logic consists of an OR operation, an AND operation, and an IMPLY operation, adding 3, 5, and 1 steps, respectively. All added together, this takes $26n-16$ steps. 


\section{Control Flow Instructions}

In order to facilitate non-sequential code execution, the RV32I instruction set provides some control flow instructions. Deviating from sequential execution is essential for implementing constructs such as loops, function calls, and conditional statements. These instructions are broadly categorized into two types: unconditional jumps (J-type), and conditional branches (B-type).

\subsection{Jumps}
In RISC-V, unconditional control flow instructions, referred to as jumps, simply write a specified value to the \gls{pc}, and store a return address ($\text{PC}+4$) to a target register. There are two such instructions in the RV32I base set: \texttt{jal} and \texttt{jalr}, which are used to implement function calls and returns.

\subsubsection{Instruction: \texttt{jal}}
\begin{figure}[h]
\centering
\includegraphics[width=1.0\textwidth]{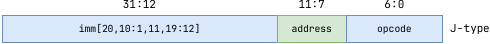}
\caption{Modified J-type encoding}
\label{fig:jtype}
\end{figure}

In the proposed instruction set, the \texttt{jal} instruction is encoded using the modified J-type format, as shown in Figure \ref{fig:jtype}. This format repurposes the standard J-type encoding, dedicating bits 30:12 to a 20-bit immediate value ($\textbf{imm}[20:1]$) that acts as a signed offset to add to the \gls{pc}. The execution of the jump involves appending a `0' to the least significant bit, sign-extending this value, and adding the result to the \gls{pc}. In short, this is expressed as the following operation:

\begin{equation}
    \text{PC} = \text{PC} + \text{SignExt}(\{\textbf{imm} [20:1], \text{`}0\text{'}\})
\end{equation}

Then, the value $\text{PC}+4$ is stored to the crossbar array at the location pointed to by the \textbf{address} field. For this, the control logic considers only the ``Start Row'' and ``Column A'' parameters from the referenced address bank entry, ignoring the remaining fields.

\subsubsection{Instruction: \texttt{jalr}}
In RISC-V, the \texttt{jalr} instruction functions very similarly to the \texttt{jal} instruction, but instead of using a large immediate value as a \gls{pc}-relative offset, it uses a register-based value. This enables the program to jump to an address calculated at runtime, which is often used for returning from a function or invoking a function pointer. The target address is obtained by adding a sign-extended 12-bit immediate to the value in a source register:
\begin{equation}
    \text{PC} = \textbf{rs1} + \text{SignExt}(\textbf{imm})
\end{equation}

In the proposed architecture, the \texttt{jalr} instruction functions the same way but sources its base address from the crossbar array rather than a register file. To support this, it uses the modified I-type format illustrated in Figure \ref{fig:itype}. The \textbf{address} field indexes the address bank to locate the base address value in the crossbar array. Unlike the previously mentioned instructions that use the modified I-type format, where the \textbf{rd} field is ignored, here it remains active. It specifies the address bank slot that points to the location in the crossbar array where the return address should be stored. Since both the source and destination are only single words, the control logic only considers the ``Start Row'' and ``Column A'' parameters of the respective addressing configurations.

The execution sequence of the \texttt{jalr} instruction is as follows:
\begin{enumerate}
    \item \textbf{Read Base Address:} The control logic reads the base address from the crossbar array at the location specified by the \textbf{address} field.
    \item \textbf{Store Return Address:} The control logic computes the value $\text{PC}+4$ and writes it to the crossbar array at the location specified by the \textbf{rd} field. The write operation uses the established two-phase SET/RESET process.
    \item \textbf{Update \gls{pc}:} Lastly, the control logic sign-extends the 12-bit immediate, adds it to the base address read in step 1, and writes the result to the PC, completing the jump.
\end{enumerate}

\subsection{Branches}

Conditional branch instructions are used to steer the control flow of a program dynamically, based on data values. These instructions compare two operands, and if a certain condition is met, update the \gls{pc} by adding a signed offset, which is used to implement loops and if-else structures. The RV32I instruction set defines six branch instructions: \texttt{beq}, 
\texttt{bne}, \texttt{blt}, \texttt{bge}, \texttt{bltu}, and \texttt{bgeu}.

\begin{figure}[h]
\centering
\includegraphics[width=1.0\textwidth]{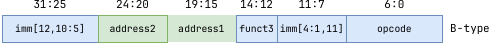}
\caption{Modified B-type encoding}
\label{fig:btype}
\end{figure}

In the proposed instruction set, these instructions are all encoded with the modified B-type format as shown in Figure \ref{fig:btype}. This format reflects the standard RISC-V B-type encoding but repurposes the source operand fields. Instead of indexing registers with \textbf{rs1} and \textbf{rs2}, these instructions use \textbf{address1} and \textbf{address2} fields, which index the address bank to locate the source operands within the crossbar array. Since branch comparisons operate only on scalar values, the control logic considers only the ``Start Row'' and ``Column A'' parameters from the referenced address bank entries.

As opposed to the other comparative instructions, which perform the operation in the crossbar array, branch comparisons are executed within the control logic itself using traditional \gls{cmos} circuitry. To do this, the control logic first reads the two source operands using sense amplifiers, and then evaluates the branch condition with a standard \gls{cmos} comparator. This is done primarily because performing 64 read operations (32 for each operand) and a \gls{cmos} comparison uses significantly less energy than all of the IMPLY operations involved in an in-memory comparison, making the proposed architecture better suited for edge computing.

According to Kvatinsky et al. \cite{Kvatinsky2014IMPLY}, reading values from the crossbar array involves pulsing the desired memristor with a voltage $\text{V}_{READ}$, and using a sense amplifier to measure the current which in turn reveals the state of the memristor. Since this operation should be non-destructive, $\text{V}_{READ}$ must be significantly smaller than the voltage $\text{V}_{SET}$, which is used in an IMPLY operation. As a consequence, the read operation draws less current, and therefore less energy than an IMPLY step.

A second advantage of this approach is that performing the comparison in-memory is needlessly destructive to the second operand. The result of the comparison is not relevant beyond the time of the branch, so there is no reason to overwrite the second operand if it can be avoided, potentially saving costly copy operations.

If the control logic determines that the branch condition is met, and the branch should be taken, it updates the \gls{pc} to the branch target address by adding the sign-extended immediate offset to the current \gls{pc}:

\begin{equation}
    \text{PC} = \text{PC} + \text{SignExt}(\{\textbf{imm}[12:1], \text{`}0\text{'}\})
\end{equation}

If the condition instead evaluates to false, the control logic takes no action, and the \gls{pc} is incremented as normal ($\text{PC} = \text{PC} + 4$).

\subsection{Architecture-Specific Instructions}

While the RV32I instruction set provides a comprehensive foundation for load-store architectures, it lacks the mechanisms required to manage the hardware of the proposed in-memory architecture. Specifically, RISC-V instructions have no concept of an address bank or the need to load addressing configurations for a large crossbar array. Furthermore, as an architecture designed for edge applications, the system requires a dedicated interface to ingest data from external sensors. To accommodate these differences, the proposed instruction set includes a small set of custom instructions not found in the RV32I standard. 

\subsubsection{Instruction: \texttt{la} (Load Address)}

The \texttt{la} instruction provides the mechanism for dynamic addressing in the proposed architecture. It reads a 32-bit data word from the crossbar array and loads it into a specific slot of the address bank. This capability is required for programs that calculate addresses at runtime and iterate through different memory locations.

\begin{figure}[h]
\centering
\includegraphics[width=1.0\textwidth]{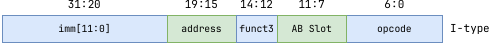}
\caption{Modified I-type encoding for address bank instructions}
\label{fig:itype_addressing}
\end{figure}

To support this operation, the \texttt{la} instruction uses the modified I-type encoding shown in Figure \ref{fig:itype_addressing}. Consistent with the addressing schema, the \textbf{rs1} field indexes the address bank to determine the source location in the crossbar array. Since the operation loads only a single word, only the ``Start Row'' and ``Column A'' fields of the referenced address are considered; the ``Num Rows'', ``Column B'', and ``Stride'' fields are ignored. Unlike other instructions that use the modified I-type encoding where the \textbf{rd} field is ignored, it is active here. It holds a direct index (0 -- 31), specifying which slot in the address bank will receive the loaded value. It should be noted that while the I-type format is selected because it conveniently provides both \textbf{rs1} and \textbf{rd} fields, the 12-bit \textbf{imm} field is entirely unused and ignored by the control logic during this operation.

The execution sequence of the \texttt{la} instruction proceeds in two phases:
\begin{enumerate}
    \item \textbf{Read Phase:} The control logic uses sense amplifiers to read the memristors where the $32$-bit source word is held.
    \item \textbf{Write Phase:} The control logic then writes this value into the address bank slot specified by \textbf{rd}. Since the address bank is physically implemented as a memristive crossbar array, this write operation follows the same procedure as the \texttt{li} instruction. In the first cycle, voltage pulses are applied to SET all bits corresponding to a logical `1'. In the second cycle, pulses are applied to RESET all bits corresponding to a logical `0'.
\end{enumerate}

\subsubsection{Instruction: \texttt{lai} (Load Address Immediate)}
The \texttt{lai} instruction acts as the mechanism for static addressing in the proposed architecture. It allows the program to load a fixed 12-bit value directly into the lower bits of a specified slot of the address bank. Static addressing is a required feature, as it provides the control logic with the initial valid addressing configurations required to target data in the crossbar array before any dynamic addresses can be calculated. 

This instruction uses the modified I-type encoding illustrated in Figure \ref{fig:itype_addressing}. In this format, the \textbf{rd} field specifies the destination slot (0--31) within the address bank. The 12-bit \textbf{imm} field contains the data to be loaded. Unlike standard arithmetic immediate instructions, the \texttt{lai} instruction does \textit{not} sign-extend the immediate value. Since the address bank holds structural configuration data rather than signed numerical values, sign extension is not necessary. Preserving the integrity of the upper 20 bits when modifying the lower 12 bits is more important. Additionally, the \textbf{rs1} field is unused and ignored, as the instruction relies solely on the immediate field for its source data.

The execution follows a direct write sequence managed by the control logic. Identical to the write phase of the \texttt{la} instruction, the operation is performed in two cycles. First, the control logic applies SET voltages to write logical `1's to the relevant bits of the target slot. In the subsequent cycle, RESET voltages are applied to write logical `0's.

\subsubsection{Instruction: \texttt{laui} (Load Address Upper Immediate)}

The \texttt{laui} instruction serves as the complement to \texttt{lai}, completing the mechanism for static addressing. While \texttt{lai} writes to the lower 12 bits of an addressing configuration, \texttt{laui} loads a fixed 20-bit value into the upper bits of a specified address bank slot.

\begin{figure}[h]
\centering
\includegraphics[width=1.0\textwidth]{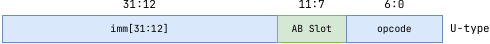}
\caption{Modified U-type encoding for \texttt{laui} instruction}
\label{fig:utype_addressing}
\end{figure}

This instruction uses the modified U-type encoding as shown in Figure \ref{fig:utype_addressing}. Identically to the \texttt{lai} instruction, the \textbf{rd} field contains a direct index indicating the target slot in the address bank. The 20-bit \textbf{imm} field contains the data to be loaded.

The execution of the write operation follows the established two-phase write sequence, writing the `1's and the `0's in two separate steps. Similarly to the \texttt{lai} instruction, the write operation is strictly confined to the upper 20 bits, leaving the lower 12 bits undisturbed. 

\subsubsection{Instruction: \texttt{lio} (Load from IO Register)}
In order to ingest data from external sources, the proposed instruction set includes the \texttt{lio} instruction. The control logic contains the IO Register, a 32-bit register that peripherals can populate, and the \texttt{lio} instruction moves the data from this register to a target location in the crossbar array.

This instruction uses the modified R-type encoding as shown in Figure \ref{fig:rtype}. The \textbf{rs1} field points to the address configuration the control logic should use as the target location. Since \texttt{lio} has only one operand, the ``Column B'' field of this addressing configuration is ignored, storing the value to ``Column A''. The ``Num Rows'' and ``Stride'' are not ignored to allow the value to be pasted to multiple rows.

\subsubsection{Instruction: \texttt{sio} (Store to IO Register)}
In order to allow data in the crossbar array to be read out externally, the instruction \texttt{sio} is proposed. It is essentially the opposite of \texttt{lio}, moving a 32-bit value from a target location in the crossbar array to the IO Register. Similarly to the \texttt{lio} instruction, it uses the modified R-type encoding as shown in Figure \ref{fig:rtype}, and it ignores the ``Column B'' field, as it has only one operand. However unlike the \texttt{lio} instruction, the ``Num Rows'' and ``Stride'' fields are ignored since only one value can be held in the IO Register.

\subsubsection{Instruction: \texttt{mv}}
In the standard RV32I instruction set, \texttt{mv} allows the value from one register to be copied to another. It is implemented as a pseudo-instruction that invokes an addition (\texttt{addi rd, rs1, 0}). While efficient in traditional \gls{cmos} logic, invoking an addition for simple data movement is extremely inefficient in a memristive crossbar. Thus, the proposed architecture implements \texttt{mv} as a native instruction using a modified I-type encoding (see Figure \ref{fig:itype}). Unlike the register-to-register transfer in standard RISC-V, the proposed implementation performs a column-to-column copy within the active rows. This instruction is necessary to preserve data that would otherwise be overwritten by another operation.

The execution of the \texttt{mv} instruction uses the algorithm shown in Table \ref{alg:move} to copy individual bits. To move an $n$-bit word, the algorithm is applied bit-wise, reusing the work memristor each time. The total number of steps and memristors required is $3n$ and $2n+1$, respectively.

\begin{table}[h]
    \centering
    \caption{IMPLY copy algorithm}
    \label{alg:move}   
    \begin{tabular}{|c|c|c|}
        \hline
        Steps & Operation  & Equivalent Logic  \\
        \hline \hline
        1 & $False(w_1, b)$ & $w_1^1=b^1=0$ \\ 
        2 & $w_1^2 = a^0 \rightarrow w_1^1$ & $w_1^2 = \overline{a}$ \\
        \rowcolor{Gray}
        3 & $b^3 = w_1^2 \rightarrow b^1$ & $b^3 = a$ \\
        \hline
    \end{tabular}
\end{table}

\subsection{System Instructions}
The RISC-V standard defines several system-level instructions necessary for managing program execution, power states, and trap handling. Examples include \texttt{wfi} (Wait for Interrupt), \texttt{ebreak} (Environment Break), and \texttt{mret} (Machine Return).

Unlike the data-processing and memory-access instructions discussed previously, these system instructions do not interact with the crossbar array, nor do they rely on serial IMPLY computation. Instead, they interface exclusively with the control logic and its internal components (such as the Control and Status Registers). Because these instructions do not entail any memristive state changes in the crossbar array or address bank, their exact hardware implementation is considered beyond the scope of this work. For the purposes of this design and the associated case study, these system instructions are treated as abstracted functional blocks within the control logic and are assumed to operate exactly as defined by the RISC-V specification.

\section{Results and Discussion} \label{section:results}

Having defined the behavior of the proposed design's instruction set in section \ref{section:Implementation}, this section now examines the practical capabilities of the design. First, the circuit level metrics of each instruction are described, then a case study is provided showcasing how the design could be used, and finally a critical discussion of the design choices made during the development of the architecture.

\subsection{Circuit-level Simulation}

\subsubsection{Simulation Set-up}
To quantify the energy consumption and verify the functional correctness of each proposed instruction, the ATOMIC \cite{seiler2024atomic} simulator was used, which uses PyLTSpice \cite{PyLTSpice} to simulate memristive algorithms at the circuit-level. The memristor model used is the VTEAM model \cite{VTEAM_Spice, Kvatinsky2015VTEAM}. The VTEAM model and ATOMIC were configured with the following parameters in Table \ref{tab:sim_params} and \ref{tab:sim_params2}.

\begin{table}[h]
    \centering
    \caption{Fitted VTEAM parameters used for simulation}
    \label{tab:sim_params}
    \begin{tabular}{|ccccccc|}
        \hline
             Parameter & $v_{off}$ & $v_{on}$ & $\alpha_{off}$ & $\alpha_{on}$ & $R_{off}$ & $R_{on}$  \\
             \hline 
             Value & 0.7V & -10mV & 3 & 3 & 1 M$\Omega$ & 10 k$\Omega$ \\
             \hline \hline
             $k_{on}$ & $k_{off}$ & $w_{off}$ & $w_{on}$ & $w_C$ & $a_{off}$ & $a_{on}$ \\
             \hline
             -0.5 nm/s & 1cm/s & 0 nm & 3 nm & 107 pm & 3 nm & 0 nm \\
             \hline
        \end{tabular}
\end{table}

\begin{table}[h]
    \centering
    \caption{IMPLY parameters used for simulation with ATOMIC}
    \label{tab:sim_params2}
        \begin{tabular}{|c|c|c|c|c|}
            \hline
            $V_{SET}$ & $V_{COND}$ & $V_{RESET}$ & $t_{pulse}$ & $R_G$ \\
            \hline \hline
            $1V$ & $0.9V$ & $-1V$ & $30ns$ & $40k\Omega$ \\
            \hline
        \end{tabular}
\end{table}

It is important to note that the VTEAM model characterizes the behavior of a memristor whose parameters are fitted to experimental data derived from the Knowm memristor--an early, discrete memristor featuring large technology nodes \cite{Kvatinsky2015VTEAM, Knowm}. Since these early discrete devices have much larger physical dimensions than integrated devices in a crossbar array, they require substantially higher drive currents to induce state changes. As such, the resulting energy metrics should be interpreted as conservative approximations. Still, they provide a reliable way of evaluating the relative costs and scaling behavior of the different instructions.

\subsubsection{Circuit-level Metrics}

As shown in Table \ref{tab:bigtable}, for each instruction in the proposed instruction set, the following metrics were deduced:
\begin{itemize}
\item \textbf{Latency:} Defined as the total number of IMPLY steps required to perform an instruction. This metric accounts for all operations performed within both the crossbar array and the address bank. Any clock cycles that may be needed by the control logic are not considered.
\item \textbf{Area:} The spatial footprint of an instruction, determined by the total number of memristors required on an active row. This includes the memristors holding the input and output operands, as well as any required work memristors.
\item \textbf{Energy Consumption:} Evaluated in nanojoules ($nJ$) using ATOMIC simulations. Because the switching energy of a memristor is asymmetrical (i.e., transitioning a memristor from an ``on'' state to an ``off'' state consumes a different amount of energy than the reverse), the reported values represent the average energy consumed across all possible input combinations.
\item \textbf{Sense Amplifier Reads (SARs):} The total number of read operations required for an instruction. This quantifies how many times the control logic must read values from the crossbar array using the sense amplifiers.
\end{itemize}

Despite the proposed architecture being an exclusively 32-bit system, adhering to the RV32I standard, the circuit-level metrics in Table \ref{tab:bigtable} are given as functions of an arbitrary bit-width of $n$. This generalized representation provides insight into the scaling behavior of bit-serial IMPLY computation. For example, it illustrates how the boolean basic and additive instructions scale with $\mathcal{O}(n)$ complexity, whereas the more complex shift operations scale with $\mathcal{O}(n\text{log}n)$ complexity. Additionally, the general $n$-bit formulation demonstrates the flexibility of the IMPLY algorithms that underlie each instruction, providing a clear expectation should future designs adapt the system for more lightweight 16-bit or higher-precision 64-bit workloads.

\begin{table}[h]
    \centering
    \caption{Circuit-level metrics for all proposed instructions (with bit-width $n$)}
    \label{tab:bigtable}
    \resizebox{\textwidth}{!}{%
    \begin{tabular}{|c|c|c|c|c|}
        \hline
        Instruction & Steps & Memristors & Avg Energy (nJ) & SARs\\
        \hline \hline
        \texttt{li} & $2$ & $0.375n$ & $0.0875n$ & $-$\\
        \texttt{lui} & $2$ & $0.625n$ & $0.1459n$ & $-$\\
        \texttt{xori} & $9n+2$ & $2n+3$ & $2.0708n$ & $-$\\
        \texttt{ori} & $3n+2$ & $2n+1$ & $1.0058n$ & $-$\\
        \texttt{andi} & $5n+2$ & $2n+1$ & $1.1634n$ & $-$\\
        \texttt{xor} & $9n$ & $2n+3$ & $1.8374n$ & $-$\\
        \texttt{or} & $3n$ & $2n+1$ & $0.7724n$ & $-$\\
        \texttt{and} & $5n$ & $2n+1$ & $0.9300n$ & $-$\\
        \texttt{addi} & $20n+2$ & $2n+4$ & $5.0590n$ & $-$\\
        \texttt{slli} & $8n\text{log}_2n -2n+4$ & $n+\text{log}_2n+2$ & $1.6248n\text{log}_2n + 0.2334\text{log}_2n - 0.5006n+0.5006$ & $-$\\
        \texttt{srli} & $8n\text{log}_2n -2n+4$ & $n+\text{log}_2n+2$ & $1.6248n\text{log}_2n + 0.2334\text{log}_2n - 0.5006n+0.5006$ & $-$\\
        \texttt{srai} & $8n\text{log}_2n+2$ & $n+\text{log}_2n+2$ & $1.6248n\text{log}_2n + 0.2334\text{log}_2n$ & $-$\\
        \texttt{add} & $20n$ & $2n+4$ & $4.8256n$ & $-$\\
        \texttt{sub} & $20n$ & $2n+4$ & $3.8525n$ & $-$\\
        \texttt{sll} & $8n\text{log}_2n -2n+2$ & $n+\text{log}_2n+2$ & $1.6248n\text{log}_2n - 0.5006n+0.5006$ & $-$\\
        \texttt{srl} & $8n\text{log}_2n -2n+2$ & $n+\text{log}_2n+2$ & $1.6248n\text{log}_2n - 0.5006n+0.5006$ & $-$\\
        \texttt{sra} & $8n\text{log}_2n$ & $n+\text{log}_2n+2$ & $1.6248n\text{log}_2n$ & $-$\\
        \texttt{auipc} & $20n+2$ & $2n+4$ & $5.2924n$ & $-$ \\
        \texttt{slti} & $26n-19$ & $2n+4$ & $5.0361n - 3.1025$ & $-$\\
        \texttt{sltiu} & $26n-15$ & $2n+4$ & $5.0361n - 2.6324$ & $-$\\
        \texttt{slt} & $26n-16$ & $2n+4$ & $4.8027n-3.1025$ & $-$\\
        \texttt{sltu} & $26n-12$ & $2n+4$ & $4.8027n - 2.6324$ & $-$\\
        \texttt{beq} & $-$ & $2n$ & $-$ & $2n$\\
        \texttt{bne} & $-$ & $2n$ & $-$ & $2n$\\
        \texttt{blt} & $-$ & $2n$ & $-$ & $2n$\\
        \texttt{bge} & $-$ & $2n$ & $-$ & $2n$\\
        \texttt{bltu} & $-$ & $2n$ & $-$ & $2n$\\
        \texttt{bgeu} & $-$ & $2n$ & $-$ & $2n$\\
        \texttt{jalr} & $2$ & $2n$ & $0.2334n$ & $n$ \\
        \texttt{jal} & $2$ & $n$ & $0.2334n$ & $-$ \\
        \texttt{la} & $2$ & $n$ & $0.2334n$ & $n$ \\
        \texttt{lai} & $2$ & $0.375n$ & $0.0875n$ & $-$\\
        \texttt{laui} & $2$ & $0.625n$ & $0.1459n$ & $-$\\
        \texttt{lio} & $2$ & $n$ & $0.2334n$ & $-$\\
        \texttt{sio} & $-$ & $n$ & $-$ & $n$\\
        \texttt{mv} & $3n$ & $2n+1$ & $0.5811n$ & $-$ \\
        \hline
    \end{tabular}
    }        
\end{table}

\subsubsection{Discussion}

As seen in Table \ref{tab:bigtable}, there is a wide range in the energy and latency costs across the proposed instruction set. Arithmetic and comparative instructions are significantly more energy-intensive and require considerably more IMPLY steps to execute than the boolean basic instructions. Aside from the \texttt{mv} instruction, all other instructions involve only direct reads from or writes to either the crossbar array or address bank. These operations are comparatively inexpensive, as any required logic (such as evaluating branch conditions) is efficiently offloaded to the CMOS-based control logic.

Since the energy consumption of the \texttt{mv} instruction is on the order of the boolean basic instructions, and it does not contribute to the processing of information, it should be avoided whenever possible. To maximize energy efficiency, code written for the proposed architecture should conform to the design's two-operand, destructive-write convention. In the case of immediate instructions, the instruction set provides some flexibility to avoid data movement. The immediate arithmetic and boolean instructions load the immediate value to ``Column B'' and subsequently overwrite it with the result. If the immediate value should persist, \texttt{lui} and \texttt{li} can be used to load an immediate value into the safe ``Column A'', and then the non-immediate variant of the desired operation can be used, overwriting the other operand. While this requires extra instructions, it provides some flexibility for a potential compiler.

Along with being the most energy-intensive operations, the arithmetic and comparative operations also incur the most latency. For example, a single \texttt{add} instruction takes 640 IMPLY cycles for standard 32-bit operands. While this bit-serial execution is dramatically slower than the single-cycle addition in traditional \gls{cmos} logic, the native parallelism of the proposed architecture redeems this limitation. Since the control logic can drive column lines across an arbitrary number of active rows simultaneously, performing an operation on an arbitrary $m$ rows introduces zero latency penalty over a single scalar operation. In effect, the ``latency per operation'' reduces by a factor of $m$. Despite the inherent slowness of serial stateful logic, the huge vectorization allows the design to remain competitive for data-parallel tasks.

Additionally, the metrics in Table \ref{tab:bigtable} further justify the decision to evaluate branch conditions inside the traditional \gls{cmos}-based control logic rather than in the crossbar array. Performing comparisons in the crossbar array is expensive; the \texttt{sltu} instruction, for instance, consumes around 160$nJ$ of energy for a 32-bit evaluation. In contrast, the \texttt{bltu} instruction costs only the minimal energy needed to read the operands via sense amplifiers, followed by a highly energy-efficient evaluation in a standard \gls{cmos} comparator. By offloading branch conditions to the control logic, the architecture saves energy and avoids a great deal of latency.

Finally, an analysis of the spatial footprint of all instructions reveals a potential optimization in the architecture. The proposed design includes 8 work memristors per row, but as can be seen in Table \ref{tab:bigtable}, the maximum needed work memristors at any given time is 4. This redundancy is not necessary, and the design could be modified to only have 4 work memristors without issue. However, if in the future, the instruction set should be expanded with more complex operations, like in-memory multiplication and division, it is possible that the need for more work memristors returns. For this reason, this design parameter should be seen as flexible.

\subsection{Case Study}
While the circuit-level metrics provide a low-level understanding of the proposed design's capabilities and resource consumption, a complete real-world application is necessary to truly evaluate the design's strengths. Edge computing devices, such as environmental sensor nodes, typically exhibit a very specific operational pattern: they sporadically collect data, perform some light local preprocessing, and spend the vast majority of their lifespan in an idle state. This profile perfectly aligns with the strengths of the proposed architecture. Since memristors are a non-volatile storage medium, the design's idle state incurs near-zero static energy loss. When active, the reduced data movement and parallelism capabilities allow it to efficiently process the accumulated data directly within the memory array.

To demonstrate the viability of the proposed \gls{isa} and architecture's addressing schema, a case study involving an intelligent temperature sensor was devised. The node is designed to measure the ambient temperature throughout the day, split into four periods (e.g., night, morning, afternoon, evening). During each period, the sensor takes eight temperature measurements, or one every 45 minutes, and it switches to a new row when a period is over. Rather than simply storing or transmitting this data, the node performs preprocessing of the data locally at the end of each day. Specifically, for each of the four periods, it calculates the average temperature and counts the number of samples that exceed a predefined period-specific threshold. To reduce the amount of instructions the design must run, this end-of-day processing is performed for all four periods in parallel, thereby reducing energy consumption.

To accomplish this, the system must be augmented with a peripheral temperature sensor and a hardware timer configured to trigger at 45-minute intervals. Using the timer, the program execution is entirely interrupt-driven. Between measurements, the architecture rests in a dormant state by invoking a Wait-for-Interrupt (\texttt{wfi}) instruction. Each time the timer expires, the sensor takes a measurement and sends an interrupt to the control logic. The design then wakes up, and executes the interrupt handler, which ingests the sample via the \texttt{lio} instruction, places it into the crossbar array, and updates the pointer to where the next sample should be stored. Only if it is the last sample of the day, the handler executes the parallel averaging and super-threshold counting before returning to the idle state. Because the node should operate continuously, the software is designed for indefinite scaling. Once the crossbar array reaches its maximum row capacity, the control logic increments its internal array ID, continuing execution on a fresh memory bank.

\subsubsection{Common Computational Patterns of the Proposed \gls{isa}}
Writing code for a destructive, two-operand architecture without traditional memory requires a new approach to assembly programming. This section explores some of the common computational patterns that arise from the capabilities and restrictions of the \gls{isa}. Only snippets of the code are shown here in order to highlight specific examples of how the proposed addressing schema interacts with the proposed \gls{isa} in ways that depart from a traditional understanding of assembly code.

\textbf{Static Addressing with Immediates:}
Static addressing occurs when the location of an operand is known at compile time. While conceptually not dissimilar to static addressing in a load-store architecture, the proposed \gls{isa} has a different format for addresses. Rather than referencing a position in a linear memory space, addresses are comprised of two indices that point to a row and column in a two-dimensional space. Furthermore, the parallelism native to the proposed \gls{isa} allow for a single instruction to target more than just one row. The code in Listing \ref{snippet:static} elucidates these concepts.
    
\begin{lstlisting}[firstnumber=124, label={snippet:static}, caption={Static addressing}]
#load threshold values to column 11 (assuming const. for each period of the year)
laui ab31 10110000000000100111
lai  ab31 111100000011			#ab31 = [11,0,4,508,3]
lui  ab31 threshold_0[31:12]
li   ab31 threshold_0[11:0]		#every fourth row, start with row4 = threshold_0

laui ab31 10110000000000101111
lai  ab31 111100000011			#ab31 = [11,0,5,508,3]
lui  ab31 threshold_1[31:12]
li   ab31 threshold_1[11:0]		#every fourth row, start with row5 = threshold_1

laui ab31 10110000000000110111
lai  ab31 111100000011			#ab31 = [11,0,6,508,3]
lui  ab31 threshold_2[31:12]
li   ab31 threshold_2[11:0]		#every fourth row, start with row6 = threshold_2

laui ab31 10110000000000111111
lai  ab31 111100000011			#ab31 = [11,0,7,508,3]
lui  ab31 threshold_3[31:12]
li   ab31 threshold_3[11:0]		#every fourth row, start with row7 = threshold_3
\end{lstlisting}
Listing \ref{snippet:static} is extracted from the initialization phase, where the address bank and crossbar array are prepared for the execution of the interrupt handler. This specific routine initializes the threshold values required for the end-of-day preprocessing. In this case study, for simplicity, it is assumed that the thresholds for each specific period (e.g., morning, afternoon) remain constant across all days.

Lines 125 and 126 use the \texttt{laui} and \texttt{lai} instructions to populate address bank slot \texttt{ab31} with an addressing configuration. The ``Column A'' field is set to 11, as the threshold values are expected in the eleventh column in the interrupt handler. The ``Column B'' field is irrelevant here, as the \texttt{lui} and \texttt{li} instructions disregard it. ``Start Row'' is set to 4, since this is the first row that sensor data is stored to, and ``Num Rows'' is set to 508, in order to fill the entire column. To facilitate the case study's requirement for four periods per day, a ``Stride'' of 3 is used, which ensures that the subsequent load-immediate instructions write the \texttt{threshold\_0} value to every fourth row (rows 4, 8, 12, etc.), populating the entire memory space allocated for that time of day. This process is repeated for the three remaining periods, with the ``Start Row'' incremented by one for each iteration. By line 143, the threshold value for every row is initialized.

\textbf{Dynamic Addressing:}
Static addressing is sufficient for initializing constants and computing on data whose location is fixed, but often this is not sufficient. Processing sequential data, such as a stream of sensor measurements, requires the ability to update memory pointers dynamically, or during program execution. In the proposed \gls{isa}, this is done by storing address configurations as data within the crossbar array, and using them as inputs to arithmetic instructions. This pattern is illustrated in Listing \ref{snippet:dyanmic}.

\newpage
\begin{lstlisting}[firstnumber=172, frame=tlr, belowskip=0pt, label={snippet:dyanmic}, caption={Dynamically managing the sample pointer}]
#ingest sensor data
la   ab31 ab0					#load sample_ptr to ab31
lio  ab31
add  ab31						#add sample to column 8
\end{lstlisting}

\vspace{-1.0em}
\centerline{\tiny \color{gray} [Lines 176--187 skipped]}

\begin{lstlisting}[firstnumber=182, frame=blr, aboveskip=0pt]
add  ab1						#iterate sample_pointer
\end{lstlisting}

In this snippet, address bank slot \texttt{ab0} acts as a ``pointer to a pointer''. It holds a static address configuration targeting [Row 0, Col 0], where a $32$-bit word representing the location of the current sample is stored. To use this ``pointer'', Line 173 executes the \texttt{la} instruction, which reads the data from [Row 0, Col 0] and writes it to address bank slot \texttt{ab31}. After Line 173 has finished, \texttt{ab31} holds a configuration where ``Column A'' points to the current sample location, ``Column B'' points to Column 8 (where all of the samples in a period are accumulated), and ``Start Row'' points to the current period's row. The subsequent \texttt{lio} and \texttt{add} instructions on Lines 174-175 use the dynamically updated address configuration to ingest the sensor data and accumulate it into the correct memory locations for that specific measurement cycle.

The most distinctive feature of dynamic addressing in the proposed \gls{isa} is how pointers are incremented. In Line 182, the \texttt{add} instruction is invoked using address bank slot \texttt{ab1}. This slot is configured with a static address configuration whose operand `A' is at [Row 0, Col 1], and operand `B' is at [Row 0, Col 0] (the location of \texttt{sample\_ptr} used in Line 173). The data in operand `A' is configured to hold a static value of [1,0,0,0,0] (interpreted as an address configuration). When the \texttt{add} instruction is completed, only the ``Column A'' field of \texttt{sample\_ptr} has been incremented, and everything else is left unchanged. It is important to note that since the address configuration is simply treated as a 32-bit integer during addition, it is difficult to check for overflow here, and thus code should be written in a manner that avoids the possibility of overflow entirely. The next time the interrupt handler triggers and Line 173 is executed, the newly incremented address is loaded into \texttt{ab31}, successfully targeting the next column in the array for the next sample. Iterating the ``Start Row'' field when a new period starts is handled analogously elsewhere in the code. This mechanism allows the architecture to handle complex data structures and iterative loops dynamically.

\textbf{Semi-Static Addressing:}
In many scenarios when writing assembly for the proposed architecture, an address configuration is neither entirely static nor entirely dynamic. A frequent requirement is to maintain a dynamically calculated field, such as the ``Start Row'', while explicitly setting other fields, such as the ``Column A'' or ``Column B''. Since addresses can be stored as data within the crossbar array, the architecture allows for this hybrid approach, using bit-wise masking to ``edit'' specific fields of a pointer without losing the dynamic context. This technique is exemplified in Listing \ref{snippet:semistatic}.

\begin{lstlisting}[firstnumber=195, label={snippet:semistatic}, caption={Setting the column fields of an existing addressing configuration}]
lui  ab6  00000000111111111111
li   ab6  111111111111          #day_ptr_modifier = [0,0,511,511,63]
and  ab6						#reset day_ptr "Column A" and "Column B" fields
lui  ab6  10001001000000000000
li   ab6  000000000000			#day_ptr_modifier = [8,9,0,0,0]
or   ab6						#day_ptr "Column A" = 8, "Column B" = 9
la   ab31 ab4
srai ab31 3                     #divide sums by 8 to obtain averages
\end{lstlisting}

In this snippet, the program prepares to calculate the average temperature for each period at the end of a day. The data pointed to by \texttt{ab4}, referred to as \texttt{day\_ptr}, already contains a dynamically updated ``Start Row'' that points to the first row of the day. It also has a ``Num Rows'' of 3 in order to target all four periods of the day. However, to perform the averaging, the addressing configuration needs to be updated to target Column 8 (where samples are accumulated) and Column 9 (the temporary location for an immediate value).

This ``semi-static'' addressing is achieved through a bitwise masking process. First, in lines 195-197, a bitmask is loaded into operand `A' of \texttt{ab6} that contains zeros in the positions corresponding to the ``Column A'' and ``Column B'' fields and ones elsewhere. Operand `B' of \texttt{ab6} points to the \texttt{day\_ptr}, so the \texttt{and} instruction resets its ``Column A'' and ``Column B'' fields. Then, in lines 198-200, a new value is loaded into operand `A' of \texttt{ab6} containing the binary values to target Column 8 and Column 9. An \texttt{or} instruction then ``pastes'' these values into the previously cleared fields of \texttt{day\_ptr}.

The final address configuration, loaded into address bank slot \texttt{ab31} in line 201, points to the dynamically determined rows, and the statically determined columns required for the subsequent shift operation. 

\subsubsection{Discussion}
The implementation of assembly code for the intelligent sensor node case study provides a good basis for discussing the broader implications of the proposed architecture and instruction set. This sub-section looks at the energy consumption in the case study and explores potential improvements to the proposed architecture and instruction set.

\textbf{Energy Analysis:}
The primary motivation for this architecture is to enable ultra-low power computation for edge devices. By counting how many of each instructions are executed every day, and using the results in Table \ref{tab:bigtable}, the total computational energy for a full day of operation was estimated to be 19.5$\mu J$. Additionally, 132 sense amplifier reads must be invoked, which consume a small amount of power as well. Missing from this estimate is the power consumed through static leakage in the \gls{cmos} control logic, but since the design is in an idle state most of the time, this is not likely to introduce a significant difference.

To contextualize this figure, it represents significantly better power performance when compared to \gls{soa} \gls{cmos} ultra-low-power microcontrollers. Commonly used sensor-node microcontrollers, such as the Texas Instruments MSP430 or Ambiq Apollo series, typically consume on the order of 10-100$mJ$ of energy per day purely due static leakage currents during deep sleep \cite{Shinde2008}. By performing all computation directly on non-volatile memristors, the proposed architecture achieves power performance orders of magnitude better than \gls{soa} ultra-low-power microcontrollers for sporadic activity cycles. Furthermore, in the proposed architecture, when a measurement is made, data only moves from the sensor to the crossbar array. In a traditional microcontroller, data has to move from the sensor, to memory, to the processor, and back to memory again. This reduction in data movement is also expected to provide significant improvements to power performance. These results suggest that the energy bottleneck for edge devices can be shifted away from data storage and processing and onto the peripheral components.

\textbf{Potential Changes to the Proposed Architecture and \gls{isa}:}
While the case study successfully demonstrates the proposed design's viability, analysis of the assembly code reveals some pain-points that could potentially be alleviated. The largest of which lies in the processes required to manipulate the address bank. As shown in Listing \ref{snippet:semistatic}, editing a single field of an address configuration requires loading full 32-bit masks and invoking successive \texttt{and} and \texttt{or} instructions. 

To reduce this, the \gls{isa} could be expanded to include a further suite of dedicated address-manipulation instructions. For example, immediate instructions could be introduced to directly overwrite individual fields of an address bank slot, such as an immediate instruction that can set either Column A or Column B. This would be advantageous, since it reduces the complexity of the code, less instructions would have to be performed, and ultimately it would reduce the energy costs associated with managing addresses. The downside is that this would stray further from the official RISC-V specification, making adapting existing compilers more difficult. Furthermore, introducing these highly specialized and architecture-specific instructions conflicts with the ``Reduced'' philosophy of RISC architectures, pushing the \gls{isa} into CISC territory.

A second potential modification concerns the global bit-width of the architecture. The current proposal uses 32-bit data words to maintain compliance with the RV32I standard. However, sensor data tends to not need such high precision. For example, the commonly used Texas Instruments TMP102 temperature sensor only has a precision of 12 bits \cite{ti_tmp102}. When operating exclusively with 12-bit measurements, most of the space in the crossbar array is wasted. 

Adopting a 16-bit version of the architecture could be advantageous for sensor-node applications. Half the bit-width theoretically halves the energy and latency of most instructions, and would allow for twice the number of samples to be stored within the same physical footprint. 16 bits is also still enough headroom to allow for the accumulation of many 12-bit samples without overflow.

Adopting the proposed architecture to operate on 16-bit data would require some adjustments to the addressing schema. First, the addressing configurations themselves would have to change. Since a 512-bit row would now hold 32 words instead of 16, the ``Column A'' and ``Column B'' fields would each require an additional bit (5 bits each). To accommodate this, the bits allotted for the ``Stride'' field would simply be reduced by 2. Alternatively, the size of the crossbar array could simply be halved, opting for only the energy efficiency and latency advantages. Additionally, since addresses remain as 32 bit values, the dynamic addressing described in Listings \ref{snippet:dyanmic} and \ref{snippet:semistatic} would no longer work, and the previously mentioned address management instructions would have to be adopted as well. While a transition to 16 bits would offer substantial performance and density benefits, this would also represent a significant departure from the RISC-V standard.

\section{Conclusions and Future Work} \label{section:conclusion}

\subsection{Conclusions}
In this thesis, we propose and evaluate a standalone, general-purpose \gls{imc} architecture tailored specifically for ultra-low-power edge devices. The rapidly expanding field of edge computing requires hardware that can operate under exceptionally strict energy constraints. Traditional \gls{cmos} microcontrollers are fundamentally limited in this domain by the von Neumann bottleneck, which wastes significant power moving data across memory buses, and by the static power leakage inherent to volatile memory. The typical design process for memristive \gls{imc} architectures in the \gls{soa} focuses predominantly on accelerators or co-processors that rely on a conventional CPU to orchestrate program flow. This approach fails to fully exploit the memristor's potential in a standalone microcontroller and lacks integration with a Turing-complete \gls{isa}. We, therefore, proposed a comprehensive architecture that unifies data storage and computation into a single physical medium using the Serial IMPLY stateful logic paradigm, eliminating the energy waste of a data bus.

We successfully mapped the standard RV32I RISC-V instruction set onto the proposed memristive architecture. Since RISC-V is fundamentally a load-store architecture constrained by a small register file, we introduced a novel address bank schema to facilitate addressing within a large two-dimensional memory space. This structure overcomes traditional register limitations and enables static, dynamic, and the novel semi-static addressing patterns. Furthermore, the address bank unlocks the native vector parallelism of the architecture through strided and multi-row configurations, allowing single instructions to execute across hundreds of rows simultaneously.

The functional completeness and ultra-low-power capabilities of the architecture were validated through both circuit-level simulations and an application case study. Using the ATOMIC simulation framework, we determined the energy, latency, and area metrics for each instruction in the proposed instruction set. These metrics were then applied to evaluate an intelligent temperature sensor node. The custom assembly code for this application demonstrated the practical viability of the supported addressing patterns and parallel data processing. The entire daily workload for the sensor node was estimated to consume only $19.5 \mu\text{J}$ of computational energy. When compared to the \gls{soa}, this represents a several-thousand-fold reduction in energy overhead, as it operates orders of magnitude below the static sleep leakage of modern ultra-low-power \gls{cmos} microcontrollers, and wastes minimal energy for data movement.

\subsection{Future Work}
The contributions presented in this thesis lay a strong foundation for the development of standalone memristive \gls{imc} architectures. Nevertheless, several promising directions for future research arise, particularly in optimizing the instruction set, expanding the software ecosystem, and the physical fabrication. Some compelling avenues for immediate optimization are the refinement of addressing instructions and experimenting with smaller bit-widths. Both avenues require balancing the trade-off between the improvements they offer and the disadvantage of further straying from the RISC-V standard.

To facilitate broader adoption, the development of a compiler backend is a crucial next step. Such a compiler must be capable of automatically managing address bank allocations and scheduling move operations to preserve variables under the two-operand destructive convention. Finally, bringing the theoretical model to real-world hardware remains a vital long-term objective. Conducting full-array SPICE simulations could also be useful to more precisely evaluate the energy use of computation in the crossbar array. In summary, this thesis presents a versatile framework for advancing standalone memristive computing, paving the way for highly efficient ultra-low-power edge architectures aligned with the needs of next-generation sensing systems.

\bibliographystyle{elsarticle-num}
\bibliography{references}

\appendix
\section{Assembly for Temperature Sensor Node} \label{appendix:code}

\begin{lstlisting}
#|================================================================|
#|============================Reference===========================|
#|================================================================|

#						    31 28 27 24 23      15 14      6  5    0
#Address Config template : [xxxx][xxxx][xxxxxxxxx][xxxxxxxxx][xxxxxx]
#						    col1  col2    row       n_rows    stride


#Address Bank:
#ab0	sample_ptr
#ab1	sample_ptr_iter
#ab2	period_ptr
#ab3	period_ptr_iter
#ab4	day_ptr
#ab5	day_ptr_iter
#ab6	day_ptr_modifier
#ab7	day_ptr_copy1	
#ab8	day_ptr_copy2
#ab9	compare_ptr
#ab10	count_superthreshold_ptr
#ab11	compare_modifier
#ab12	count_superthreshold_modifier
#ab13	
#ab14	
#ab15	
#ab16	sample_count 
#ab17	max_samples
#ab18	period_count
#ab19	max_periods
#ab20	day_count
#ab21	max_days
#ab22	count_iter
#ab23	
#ab24	
#ab25	
#ab26	
#ab27	
#ab28	
#ab29	temp
#ab30	temp
#ab31	temp


#|================================================================|
#|=======================Initialization Code======================|
#|================================================================|

#set up address bank
laui ab0  00000000000000000000
lai  ab0  000000000000			#ab0 = [0,0,0,0,0] (address of sample_ptr)
laui ab1  00010000000000000000
lai  ab1  000000000000			#ab1 = [1,0,0,0,0] (address of sample_ptr_iter)
laui ab2  00100000000000000000
lai  ab2  000000000000			#ab2 = [2,0,0,0,0] (address of period_ptr)
laui ab3  00110010000000000000
lai  ab3  000000000000			#ab3 = [3,2,0,0,0] (address of period_ptr_iter)
laui ab4  01000000000000000000
lai  ab4  000000000000			#ab4 = [4,0,0,0,0] (address of day_ptr)
laui ab5  01010100000000000000
lai  ab5  000000000000			#ab5 = [5,4,0,0,0] (address of day_ptr_iter)
laui ab6  01100000000000000000
lai  ab6  000000000000			#ab6 = [6,4,0,0,0] (address of day_ptr_modifier)
laui ab7  01000111000000000000
lai  ab7  000000000000			#ab7 = [4,7,0,0,0] (day_ptr_copy1)
laui ab8  01001000000000000000
lai  ab8  000000000000			#ab8 = [4,8,0,0,0] (day_ptr_copy2)
laui ab9  01110000000000000000
lai  ab9  000000000000			#ab9 = [7,0,0,0,0] (address of compare_ptr)
laui ab10 10000000000000000000
lai  ab10 000000000000			#ab10 = [8,0,0,0,0] (address of count_superthreshold_ptr)
laui ab11 10010111000000000000
lai  ab11 000000000000			#ab11 = [9,7,0,0,0] (address of compare_modifier)
laui ab12 10101000000000000000
lai  ab12 000000000000			#ab12 = [10,8,0,0,0] (address of count_superthreshold_modifier)

laui ab16 00000000000000001000
lai  ab16 000000000000			#ab16 = [0,0,1,0,0] (address of sample_count)
laui ab17 00010000000000001000
lai  ab17 000000000000			#ab17 = [1,0,1,0,0] (address of max_samples)
laui ab18 00100000000000001000
lai  ab18 000000000000			#ab18 = [2,0,1,0,0] (address of period_count)
laui ab19 00110000000000001000
lai  ab19 000000000000			#ab19 = [3,0,1,0,0] (address of max_periods)
laui ab20 01000000000000001000
lai  ab20 000000000000			#ab20 = [4,0,1,0,0] (address of day_count)
laui ab21 01010000000000001000
lai  ab21 000000000000			#ab21 = [5,0,1,0,0] (address of max_days)
laui ab22 01100000000000001000
lai  ab22 000000000000			#ab22 = [6,0,1,0,0] (address of sample_count_iter)
laui ab23 01100010000000001000
lai  ab23 000000000000			#ab23 = [6,2,1,0,0] (period_count_iter)
laui ab24 01100100000000001000
lai  ab24 000000000000			#ab24 = [6,4,1,0,0] (day_count_iter)
#load initial values
lui  ab0  00001000000000100000
li   ab0  000000000000			#sample_ptr = [0,8,4,0,0]
lui  ab1  00010000000000000000
li   ab1  000000000000			#sample_ptr_iter = [1,0,0,0,0]
lui  ab2  00001000000000100000
li   ab2  000000000000			#period_ptr = [0,8,4,0,0]
lui  ab3  00000000000000001000
li   ab3  000000000000			#period_ptr_iter = [0,0,1,0,0]
lui  ab4  00000000000000100000
li   ab4  000011000000			#day_ptr = [0,0,4,3,0]
lui  ab5  00000000000000100000
li   ab5  000000000000			#day_ptr_iter = [0,0,4,0,0]

lui  ab16 0
li   ab16 0						#sample_count = 0
lui  ab17 0
li   ab17 7						#max_samples = 7
lui  ab18 0
li   ab18 0						#period_count = 0
lui  ab19 0
li   ab19 3						#max_periods = 3
lui  ab20 0
li   ab20 0						#day_count = 0
lui  ab21 0
li   ab21 127					#max_days = 127
lui  ab22 0
li   ab22 1						#count_iter = 1

#load threshold values to column 11 (assuming constant for each period throughout the year)
laui ab31 10110000000000100111
lai  ab31 111100000011			#ab31 = [11,0,4,508,3]
lui  ab31 threshold_0[31:12]
li   ab31 threshold_0[11:0]		#every fourth row, starting with row 4 = threshold_0

laui ab31 10110000000000101111
lai  ab31 111100000011			#ab31 = [11,0,5,508,3]
lui  ab31 threshold_1[31:12]
li   ab31 threshold_1[11:0]		#every fourth row, starting with row 5 = threshold_1

laui ab31 10110000000000110111
lai  ab31 111100000011			#ab31 = [11,0,6,508,3]
lui  ab31 threshold_2[31:12]
li   ab31 threshold_2[11:0]		#every fourth row, starting with row 6 = threshold_2

laui ab31 10110000000000111111
lai  ab31 111100000011			#ab31 = [11,0,7,508,3]
lui  ab31 threshold_3[31:12]
li   ab31 threshold_3[11:0]		#every fourth row, starting with row 7 = threshold_3

#reset column 8 (where samples are accumulated)
laui ab31 10000000000000100111
lai	 ab31 111100000000			#ab31 = [8,0,4,508,0]
lui  ab31 0
li   ab31 0

#reset column 10 (where super-threshold values will be counted)
laui ab31 10100000000000100111
lai	 ab31 111100000000			#ab31 = [10,0,4,508,0]
lui  ab31 0
li   ab31 0
			
#loop and wait for interrupt
wait:
wfi
beq  ab20 ab21 end				#if max_days reached, end program and switch to new array

loop:
beq  ab0  ab0  wait

end:
nxt_array

#|================================================================|
#|========================Interrupt Handler=======================|
#|================================================================|

#ingest sensor data
la   ab31 ab0					#load sample_ptr to ab31
lio  ab31
add  ab31						#add sample to column 8

#if max_samples reached, handle new period behavior
beq  ab16 ab17 sample_if

#usually just iterate sample_count and sample_ptr
add  ab22						#iterate sample_count
add  ab1						#iterate sample_pointer
beq  ab0  ab0  sample_end

#if max_periods reached, handle new day behavior
sample_if:
beq  ab18 ab19 period_if

#usually just iterate period_count
add  ab23
beq  ab0  ab0  period_end

#compute averages and count super-threshold values
period_if:
lui  ab6  00000000111111111111
li   ab6  111111111111			#day_ptr_modifier = [0,0,511,511,63]
and  ab6						#reset day_ptr "Column A" and "Column B" fields
lui  ab6  10001001000000000000
li   ab6  000000000000			#day_ptr_modifier = [8,9,0,0,0]
or   ab6						#day_ptr "Column A" = 8, "Column B" = 9
la   ab31 ab4

#divide sums by 8 to obtain averages
srai ab31 3

#set up address bank for super-threshold counting
mv   ab7						#copy day_ptr to compare_ptr
mv   ab8						#copy day_ptr to count_superthreshold_ptr
lui  ab11 00000000111111111111
li   ab11 111111111111			#compare_modifier = [0,0,511,511,63]
lui  ab12 00000000111111111111
li   ab12 111111111111			#count_superthreshold_modifier = [0,0,511,511,63]
and  ab11						#reset compare_ptr "Column A" and "Column B"
and  ab12						#reset count_superthreshold_ptr "Column A" and "Column B"
lui  ab11 10111001000000000000
li   ab11 000000000000			#compare_modifier = [11,9,0,0,0]
lui  ab12 10011010000000000000
li   ab12 000000000000			#count_superthreshold_modifier = [9,10,0,0,0]
or   ab11						#compare_ptr "Column A" = 11, "Column B" = 9
or   ab12						#count_superthreshold_ptr "Column A" = 9, "Column B" = 10
la   ab29 ab9
la   ab30 ab10

#check samples 0-7 unrolled to avoid expensive increments
#samples 0
lui  ab6 00001111111111111111
and  ab6						#day_ptr "Column A" = 0 "Column B" = 9
la   ab31 ab4
mv   ab31
slt  ab29						#check if sample exceeds threshold
add  ab30						#increment count if sample exceeded threshold

#samples 1
lui  ab6 00001111111111111111
and  ab6						#reset day_ptr "Column A"
lui  ab6 00010000000000000000
or   ab6						#day_ptr "Column A" = 1 "Column B" = 9
la   ab31 ab4
mv   ab31
slt  ab29						#check if sample exceeds threshold
add  ab30						#increment count if sample exceeded threshold

#samples 2
lui  ab6 00001111111111111111
and  ab6						#reset day_ptr "Column A"
lui  ab6 00100000000000000000
or   ab6						#day_ptr "Column A" = 2 "Column B" = 9
la   ab31 ab4
mv   ab31
slt  ab29						#check if sample exceeds threshold
add  ab30						#increment count if sample exceeded threshold

#samples 3
lui  ab6 00001111111111111111
and  ab6						#reset day_ptr "Column A"
lui  ab6 00110000000000000000
or   ab6						#day_ptr "Column A" = 3 "Column B" = 9
la   ab31 ab4
mv   ab31
slt  ab29						#check if sample exceeds threshold
add  ab30						#increment count if sample exceeded threshold

#samples 4
lui  ab6 00001111111111111111
and  ab6						#reset day_ptr "Column A"
lui  ab6 01000000000000000000
or   ab6						#day_ptr "Column A" = 4 "Column B" = 9
la   ab31 ab4
mv   ab31
slt  ab29						#check if sample exceeds threshold
add  ab30						#increment count if sample exceeded threshold

#samples 5
lui  ab6 00001111111111111111
and  ab6						#reset day_ptr "Column A"
lui  ab6 01010000000000000000
or   ab6						#day_ptr "Column A" = 5 "Column B" = 9
la   ab31 ab4
mv   ab31
slt  ab29						#check if sample exceeds threshold
add  ab30						#increment count if sample exceeded threshold

#samples 6
lui  ab6 01101111111111111111
and  ab6						#reset day_ptr "Column A"
lui  ab6 00010000000000000000
or   ab6						#day_ptr "Column A" = 6 "Column B" = 9
la   ab31 ab4
mv   ab31
slt  ab29						#check if sample exceeds threshold
add  ab30						#increment count if sample exceeded threshold

#samples 7
lui  ab6 01111111111111111111
and  ab6						#reset day_ptr "Column A"
lui  ab6 00010000000000000000
or   ab6						#day_ptr "Column A" = 7 "Column B" = 9
la   ab31 ab4
mv   ab31
slt  ab29						#check if sample exceeds threshold
add  ab30						#increment count if sample exceeded threshold

#set up for next day
li   ab18 0						#period_count = 0
add  ab24						#increment day_count
lui  ab6 00001111111111111111
and  ab6						#reset day_ptr "Column A"
add  ab5						#increment day_ptr

#set up for next period
period_end:
add  ab3						#iterate period_ptr
laui ab31 00100000000000000000
lai  ab31 000000000000			#ab31 = [2,0,0,0,0]
mv   ab31						#sample_ptr = period_ptr
li   ab16 0						#sample_count = 0

sample_end:
mret
\end{lstlisting}

\end{document}